\documentclass[aps,multicolumn]{revtex4}

\usepackage{graphicx}

\begin{document}

\title{Low temperature electron transfer in strongly condensed phase}

\author{Joachim Ankerhold}
\email[]{ankerhold@physik.uni-freiburg.de}
\author{Hartwig Lehle}
\affiliation{ Physikalisches Institut, Albert-Ludwigs-Universit{\"a}t
Freiburg, Hermann-Herder-Stra{\ss}e 3, 79104 Freiburg, Germany}

\date{\today}

\begin{abstract}
Electron transfer coupled to a collective vibronic degree of freedom
is studied in strongly condensed phase and at lower temperatures where
quantum fluctuations are essential. Based on an exact representation of the
reduced density matrix of the electronic+reaction coordinate compound
in terms of path integrals, recent findings on the overdamped limit in
quantum dissipative systems are employed. This allows to give for
the first time a consistent
generalization of the well-known Zusman equations to the quantum
domain. Detailed conditions for the range of validity are
specified. Using the Wigner transform these results are also extended to the 
quantum dynamics in full phase space. As an important application
electronic transfer rates are derived that comprise adiabatic and
nonadiabatic processes in the low temperature regime including nuclear
tunneling. Accurate agreement with precise quantum Monte Carlo data is
observed.

\end{abstract}

\maketitle

\section{Introduction}\label{intro}
 
Electron transfer (ET) in complex environments is ubiquitous in physics,
chemistry, and biology, the most prominent example being
photosynthesis \cite{marcus,jortner}. In the last decade with the
advent of pump-probe 
techniques the detection of fast electron transfer dynamics in
molecular systems became accessible and meanwhile even coherent vibronic
motion interacting with  transfer processes has been observed in a
variety of condensed phase structures
\cite{vos91}--\cite{eng99}. Experimentally, the transfer 
dynamics depends very sensitively on a bunch of parameters such as
 temperature, electronic 
couplings, structure of the environmental degrees of freedom, to name
but a few.

Despite this complexity theoretical foundations to describe these systems
have already been laid in the 80ties with the pioneering work by Marcus et
al.\ \cite{marcus}. It turned out that 
qualitatively two dynamical
domains must be distinguished: For vibronic dynamics fast compared to
the bare ET, characteristic for long-distance ET, one speaks of nonadiabatic
reactions, in the opposite case of very slow vibronic motion, realized
e.g.\ in mixed-valence compounds, the reaction is said to be
adiabatic. Tremendous
progress towards a deeper understanding of this picture
has been made for the
archetypical case of  donor-acceptor systems, especially based on 
descriptions that include the coupling to the residual molecular degrees
of freedom  also at lower temperatures. The common strategy 
starts from system+reservoir models and  considers the reduced
dynamics of the system (relevant degrees of freedom) only
\cite{weiss}. In this context a standard 
treatment has been Redfield theory \cite{red1}--\cite{wolf96}, but also path
integral approaches have been exploited \cite{weiss}, the latter one
particularly suitable for strongly condensed phase systems.  
Within these formulations transfer rates have been successfully
calculated in the nonadiabatic regime over the whole temperature range
by invoking golden rule techniques. Likewise, rate calculations for adiabatic
reactions  at sufficiently high temperatures exploited
well-established tools for activated rare events 
as e.g.\ Kramers' flux over population method. These quite different
approaches reflect the different physical processes that control the
transfer, namely, in the former case the electronic 
coupling between diabatic donor/acceptor states,  in the latter  one the 
sluggish activated bath motion on the lower adiabatic surface.
 To find a formulation that interpolates, at least partially,
between  these 
distinct dynamical ranges has been one of the fundamental problems in
ET theory. 

A crucial step forward into this direction has been first achieved by
Zusman \cite{zusman} 
and then later within the path integral 
approach by Garg et al.\ \cite{garg}. For sufficiently high temperatures and 
moderately fast to slow bath modes they derived equations of motion
for the electronic dynamics interacting with a damped collective degree of
freedom, coined the reaction coordinate (RC). 
In essence,  a classical  overdamped RC motion
(Smoluchowski limit) is coupled to the Heisenberg equations of the
electronic degree of freedom.
These so-called Zusman equations
(ZE) not only opened the door to study the relaxation dynamics of the
compound in detail \cite{cukier}--\cite{thoss}, but also to derive an explicit
expression for the 
relaxation rate \cite{marcus,zusman}. The Marcus/Zusman rate comprises
to some extent 
nonadiabatic and 
adiabatic dynamics as well, and reduces to the
high temperature golden rule result for small electronic coupling and
to the corresponding activated Kramers rate for large coupling. A
thorough discussion 
of the various  rate expressions and their performance  in
comparison with precise quantum Monte Carlo results has been given in
\cite{lothar}. 

The only range where so far no analytical description has been
available includes moderately fast to slow bath modes as in the ZE,
but at low temperatures. For a RC this domain corresponds to
overdamped {\em quantum}
dynamics which has been analyzed only recently in
\cite{ankerprl,ankerepl}: There it has been shown that the position
probability 
distribution of a RC obeys a Smoluchowski type of equation also at
lower temperatures with a substantial impact of  quantum fluctuations;
for
high temperatures the quantum Smoluchowski equation reduces to the
known classical one. The basic idea of the present
work is now to investigate to what extent these findings can be adapted 
 to the ET problem and to arrive at a generalization of the
ZE to the quantum domain. For that purpose the 
 path integral representation must be utilized meaning here to start
 with the exact 
density matrix of the reduced dynamics for the
electron+RC. Fortunately, the corresponding rather involved
calculation has already 
been carried out in \cite{lucke}. In the sequel, we directly built on these
achievements 
and derive in a kind of semiclassical analysis from the exact
expression approximate generalized Zusman equations (GZE). One particular
advantage of the approach is that the approximations are well
controlled and detailed conditions for the range of validity  of the
GZE can be given.

The article is organized as follows. We introduce the model and
collect the main results for 
the reduced dynamics in Sec.~\ref{dynamics}. In Sec.~\ref{qzusman} we
start with a discussion of the  quantum Smoluchowski
limit to proceed with the derivation of the generalized Zusman
equations from the exact path integral expression.  An extension
to the full RC phase space in terms of Wigner functions is achieved in
Sec.~\ref{fokker}. As an application we derive in Sec.~\ref{rate} an
analytic expression for the ET rate which generalizes the classical
Zusman result to lower temperatures and compares well with numerically
exact quantum Monte Carlo data. Conclusions are given in Sec.~\ref{conclu}.

\section{Model Hamiltonian And Reduced Density Matrix}\label{dynamics}

We consider intermolecular ET between localized
electronic states, called donor and acceptor henceforth, in presence
of a vibronic environment.
Basically two models have been developed to describe the dynamics of
this compound, namely, the spin-boson model \cite{leggett87} and the reaction
coordinate (RC) model \cite{garg}. Within the reduced
subspace of the electronic dynamics both models are completely
equivalent, but they differ in the representation
of the environmental degrees of freedom and thus allow to focus on
different aspects of the transfer process. In the spin-boson model the
electronic two state system is linearly coupled to a bath of harmonic
oscillators, while in the reaction coordinate picture the electronic
system interacts with a collective bath degree of freedom (reaction
coordinate) which is embedded in a harmonic environment. The latter
modeling allows to study explicitly the combined dynamics of electronic
and collective vibronic motion and particularly the influence of
vibronic wave packet motion onto the transfer process.

\subsection{Electron Transfer Model}\label{etmodel}

The corresponding reaction coordinate Hamiltonian consists of four
parts, namely,
\begin{equation}
H  =   H_{EL} + H_{RC} + H_I + H_B \label{et1}
\end{equation}
where
\begin{equation}
H_{EL}
     =  -\frac{\hbar\epsilon}{2} \sigma_z
        -  \frac{\hbar\Delta}{2} \sigma_x \label{et2}
\end{equation}
represents the bare electronic two-state system as an artificial
spin-$\frac{1}{2}$ system with donor $|-\rangle$ and acceptor
$|+\rangle$,  a bias $\hbar \epsilon$, and an ET coupling $\hbar\Delta$ which
is assumed to be independent of the nuclear degrees of freedom. The
second part
\begin{equation}
H_{RC}
     =      \frac{p^2}{2m}
        + \frac{m\omega_0^2}{2}q^2 -c_0 q \sigma_z\label{et3}
\end{equation}
describes a harmonic collective vibronic degree of freedom (reaction
coordinate, RC) linearly coupled to the electronic system and interacting
with a heat bath of residual molecular degrees of freedom
\begin{eqnarray}
H_I
    & =&     -q \sum c_j y_j
        + q^2 \sum\frac{c_j^2}{2m_j\omega_j^2} \\
H_B
    & = &   \sum \left\{ \frac{p_j^2}{2m_j}
        + \frac{m_j  \omega_j}{2}
        y_j^2   \right\}\, .\label{et4}
\end{eqnarray}
This way the environmental oscillators couple to the transfer system
only indirectly via the reaction coordinate $q$ (secondary bath). Of
course, one could
transform both $q$ and $\{y_j\}$ to normal modes which leads  to
the spin-boson model \cite{garg,chandler-lfgt}. Here, however, we are
interested in the 
combined dynamics of spin+reaction coordinate and thus need to
calculate the time-dependent reduced density matrix $\rho_{\sigma,
  \sigma'}(q,q',t)$ obtained by integrating out the secondary bath.
This can be done exactly as we have shown in \cite{lucke} and we will
 summarize the main findings in the next section.

Before we do so, let us address the initial condition for the time
evolution which turns out to be a crucial point, especially in the
strongly condensed phase limit. We imagine a donor state as
being excited from some ground state (dark state) at time $t=0$ (see
fig.~1). The
excitation pulse is assumed to be sufficiently short so that there are no
correlations between the electron and its surrounding at $t=0$ meaning
that the total initial density matrix $W_0$ factorizes into a direct product
of a projector onto a pure donor state and a density operator
$\rho_0$ of RC and secondary bath, i.e.,
\begin{equation}
 W_0=|-\rangle\langle -|\, \otimes\, \rho_0\, .  \label{et5}
\end{equation}
The generalization to  other electronic initial states is
straightforward (see below).
For $\rho_0$ two qualitatively different cases have to be
distinguished. In the first one RC and secondary bath factorize
corresponding to an uncorrelated initial state. This ``preparation''
is well-known from Feynman-Vernon theory \cite{feynman-vernon} and is
clearly reasonable if 
RC and bath are only weakly coupled. In case of strong coupling, however,
 it is more realistic to assume a correlated
 initial state \cite{report} 
\begin{equation}
\rho_0^{\rm cor}(q,q',\{y_n\},\{y_n'\}) = \lambda(q,q')\,
\tilde{\rho}_\beta(q,\{y_n\},q',\{y_n'\})\label{et6}
\end{equation}
with the equilibrium density matrix of  RC and secondary bath in the
donor state 
\begin{equation}
\tilde{\rho}_\beta(q,\{y_n\},q',\{y_n'\})
        =   \frac{1}{Z_{\rm B,RC}}\, \langle q,\{y_n\} |
            \exp \left[ -\beta \left( H_B +H_I
            + \frac{p^2}{2m}
            +\frac{m\omega_0^2}{2}q^2 +c_0q \right)\right]
            |q',\{y_n'\}\rangle\label{et7}
\end{equation}
at inverse temperature $\beta=1/k_{\rm B} T$ and with $Z_{\rm B,RC}$
the partition function of the bath+RC compound. Deviations from
equilibrium in the RC are created by a preparation
function $\lambda(q,q')$. In
fact,
only starting with correlated initial states guarantees to regain
from the exact quantum dynamics of the RC in the high
temperature/strong damping limit the 
so-called Smoluchowski equation for its position probability
distribution \cite{hanggicorr,pechukas}.

\subsection{Reduced Dynamics}\label{reduced}

Exploiting Feynman's path integral formalism for the time propagation of
the reduced density
\begin{equation}
\rho(t)={\rm Tr}_B \{\exp(-iHt/\hbar)W_0 \exp(iHt/\hbar)\}\label{red1}
\end{equation}
the harmonic degrees of freedom of the secondary bath can be
eliminated exactly \cite{cal-leg1,cal-leg2}. This leads to the exact
expression 
\begin{equation}
\rho_{\sigma_f \sigma_f'}(q_f,q_f', t)
     =   \int_{\sigma(0)=-1}^{\sigma(t)=\sigma_f}{\cal D}\sigma
        \int_{\sigma'(0)=-1}^{\sigma'(t)=\sigma_f'}
              {\cal D}\sigma'
        \, e^{\frac{i}{\hbar} (S_{EL}[\sigma]-S_{EL}[\sigma'])} \
    \tilde{\rho}(q_f,q_f',t; [\sigma],
    [\sigma'])\, .\label{red2}
\end{equation}
Here, forward and backward electronic path $\sigma(s), \sigma'(s)$
connect in time $t$ the initial points $\sigma(0)=-1, \sigma'(0)=-1$ with
$\sigma_f, \sigma_f'$ where each path is weighted with the bare
electronic action $S_{\rm EL}$ and the density matrix of the damped
RC. The latter takes the form
\begin{equation}
\tilde{\rho}(q_f,q_f',t;[\sigma],[\sigma'])
     =      \frac{1}{Z}\int_{-\infty}^\infty dq_i \, dq_i'\
               \lambda(q_i,q_i')
               \ J(q_f,q_f',t,q_i,q_i';[\sigma],[\sigma'])
    \label{red3}
\end{equation}
where the propagating function $J(\cdot)$ is a threefold path
integral (two in real time, one in imaginary time) over the RC degree
of freedom only. The forward and backward real time paths $q(s), q'(s)$
run in time $t$ for
given electronic paths $\sigma(s), \sigma'(s)$ from $q_i,q_i'$ to
fixed end points $q_f, q_f'$, while the imaginary time paths
$\bar{q}(\tau)$ connect $q_i$ with $q_i'$ in the interval
$\hbar\beta$. The contribution of each path is weighted according to
its effective action $\Sigma_{\rm RC}$, i.e.,
\begin{equation}
\Sigma_{\rm RC}[\bar{q},q,q';\sigma,\sigma']=i \bar{S}_{\rm
  RC}[\bar{q}]+S_{\rm
  RC}[q;\sigma,\sigma']- S_{\rm
  RC}[q';\sigma,\sigma']+\phi[\bar{q},q,q']\, .\label{red4}
\end{equation}
It consists
of the bare actions of the RC system according to $H_{\rm RC}\ $
({\ref{et3}) in real and imaginary time and
  the so-called influence functional describing the effective influence of the
  secondary bath onto the RC.
The real time paths encode the dynamics of the RC system and
the imaginary time paths specify its initial preparation, namely,
\begin{equation}
\tilde{\rho}(q_f,q_f',0;\sigma_i,\sigma_i')=\frac{1}{Z}
               \lambda(q_f,q_f')\
               \tilde{\rho}_\beta(q_f,q_f')\label{red5}
\end{equation}
with  $\tilde{\rho}_\beta(q,q')=\langle
q|{\rm Tr}_{\rm B} \exp[-\beta(H_{\rm RC}+H_{\rm B}+H_{\rm
    I})]|q'\rangle|_{ \sigma_z=-1}$ the reduced equilibrium of the RC
in the donor.
The normalization in (\ref{red4}) and (\ref{red5}) is chosen such
that for vanishing RC-bath coupling the influence functional reduces
to 1, i.e. $Z=Z_{\rm B,RC}/Z_{\rm B}$ where $Z_{\rm B}$ is the
partition function of the secondary bath alone.
We omit here the
explicit form of the effective action and refer to
\cite{lucke,report} for further details. Note that since electronic and
RC-system are initially uncorrelated the Euclidian action $\bar{S}_{\rm
  RC}[\bar{q}]$ in
(\ref{red4}) is independent of the electronic paths. The
influence of the harmonic bath onto the RC described by the influence
functional $\phi[\bar{q},q,q']$ is completely covered by
the autocorrelation
function of the fluctuating force $\sum c_j y_j$ [see (\ref{et3})]
\begin{eqnarray}
K(\theta) &=& \frac{1}{\hbar}\, \left\langle\left[\sum c_j y_j(\theta)\right]\left[\sum c_j y_j(0)\right]\right\rangle_\beta\nonumber\\
 &=&    \int_0^\infty \frac{d\omega}{\pi} I(\omega)
    \frac{\cosh
    [\omega(\hbar\beta/2-i\theta)]}{\sinh(\omega\hbar\beta/2)}\label{red6}
\end{eqnarray}
where $\theta=s-i\tau$, $0\leq s\leq t$, $0\leq\tau\leq\hbar\beta$. All
relevant bath properties  are contained in the spectral density
$I(\omega)$ of the bath oscillators which, as a classical quantity, can
be obtained from molecular dynamics simulations. For real times
$K(s)=K'(s)+i K''(s)$
is directly related to the macroscopic damping kernel
\begin{equation}
\gamma(s)   =   \frac{2}{m}\int_0^\infty \frac{d\omega}{\pi}
            \frac{I(\omega)}{\omega}\cos(\omega s)\label{red7}
\end{equation}
via $K''(s)=(m/2)d\gamma(s)/ds$ and $K'(s)\to
m\gamma(s)/\hbar\beta$ in the classical limit $\omega_c\hbar\beta\ll 1$
with a typical bath frequency  $\omega_c$ usually taken as the
bath cut-off frequency. The damping kernel produces interactions
along the RC paths that are non-local in time even in cases where
$\omega_c$ is very large and thus, classically, the corresponding bath
memory time very
short. This is due to quantum fluctuations that appear on a time scale
$\hbar\beta$ and are responsible for non-Markovian behavior at lower
temperatures.

\subsection{Exact Reduced Density Matrix for the Reaction
  Coordinate}\label{redRC}

For given electronic paths the dissipative dynamics of the RC
looks like the dynamics of a damped harmonic oscillator in presence of
force fields \cite{report}. The crucial difference to the case of an
external force, 
however,
is the fact that electronic forward and backward paths are not
identical, but rather are dynamical degrees of freedom as well.
The corresponding somewhat involved calculation to obtain the reduced
density of the RC has been presented in
\cite{lucke}.
Eventually,  one ends up with an exact expression for the
reduced dynamics of the RC
\begin{equation}
\tilde{\rho}(x_f, r_f,t;[\chi],[\eta])
    =\frac{1}{N(t)}\,   \int dx_i\,dr_i\,\lambda(x_i,r_i)\
        \exp \left[\frac{i}{\hbar}
        \Sigma_{\rm ma}(x_f,r_f,t,x_i,r_i;[\chi],[\eta])
        \right]\label{rede1}
\end{equation}
where for convenience we introduced sum and difference paths for the
spin
\begin{equation}
\chi=(\sigma-\sigma')/2\, ,\ \ \ \eta=(\sigma+\sigma')/2\label{rede2}
\end{equation}
and sum and difference coordinates for the RC
\begin{equation}
x_\alpha=q_\alpha-q'_\alpha\, ,\ \ \ r_\alpha=(q_\alpha+q'_\alpha)/2\,
 ,\ \ \ \ \ \ \alpha=i,f\, .\label{rede3}
\end{equation}
The exponent $\Sigma_{\rm ma}$ is the effective action evaluated at
the minimal action paths and $N(t)$ is the normalization. The explicit
form of both is not very illuminating (for details see
  \cite{lucke}); $\Sigma_{\rm ma}$ is a quadratic
  function of the RC-coordinates with time dependent coefficients that
  are functionals of $\chi(s)$ and $\eta(s)$, while $N(t)$ depends
  only on time. For given spin paths their
  time dependence can
  be completely expressed in terms of the real and imaginary parts of the
  autocorrelation function of a damped harmonic oscillator
  $\langle q(t)q\rangle$.

The exact dynamics of the electronic+RC system in presence of the
residual vibronic degrees of freedom is now given by (\ref{red2})
together with (\ref{rede1}). If one would integrate over the diagonal
part $\tilde{\rho}(0,r_f,t;[\chi],[\eta])$ too by putting
$\lambda=1$, one would recover the
exact reduced path integral expression for the reduced spin-boson
dynamics. Here, however, we are able to study the relaxation dynamics
of the RC starting with nonequilibrium initial
states explicitly. For that purpose
what remains to be done, is an evaluation of the spin
path integrals which is not possible analytically. The
fundamental complication lies in the self-interactions, non-local in
time, and mediated by the secondary bath. As a consequence, the dynamics of
the density matrix $\rho_{\sigma_f,\sigma_f'}(x_f,r_f,t)$ can in general not be
determined from simple equations of motions so that one has to work
within the path integral representation. Progress can be made by
applying either numerical methods like Quantum Monte Carlo techniques
\cite{lothar} 
or perturbative approaches like e.g.\ NIBA or golden rule \cite{weiss}.
As was shown by Zusman \cite{zusman} and then in detail by Garg et
al.\ \cite{garg} 
particularly the high
temperature/strong damping limit allows for an effectively Markovian
description.
There, only the
diagonal part
\begin{equation}
P_{\sigma_f,\sigma_f'}(q_f,t)=\rho_{\sigma_f,\sigma_f'}(x_f=0,r_f=q_f,t)
\label{rede3b}
\end{equation}
 is
relevant and its time evolution is determined by a set of equations of
motion where the bare electronic dynamics is coupled to a classical
overdamped RC
motion (Smoluchowski limit). Our first goal is to derive {\it
  generalized Zusman equations} which are also valid in the low
temperature range.

\section{Generalized Zusman Equations}\label{qzusman}

While the strong friction limit is well-known in classical mechanics
\cite{risken}, 
corresponding quantum systems have been analyzed only recently
\cite{ankerprl}. Hence,
in this section we first summarize some main findings which then serve
as a basis for the treatment of the ET system with overdamped RC.

\subsection{Classical and Quantum Smoluchowski Limit}\label{smolu}

The strong friction limit in classical physics, known as the
 Smoluchowski limit,  is characterized by a
separation of time scales between equilibration of momentum and
equilibration of position. This allows to adiabatically eliminate the
momentum from the phase space Fokker-Planck equation and to gain a
simple time evolution equation for the position distribution, the
Smoluchowski equation \cite{skinner}.
In case of a harmonic oscillator with frequency $\omega_0$ the basic
condition then is $\gamma\gg \omega_0$ where $\gamma$ denotes a
typical damping strength. This is defined
as
\begin{equation}
\gamma\equiv \hat{\gamma}(0)=\lim_{\omega\to
  0}\frac{I(\omega)}{m\omega}\label{smol1}
\end{equation}
with $\hat{\gamma}(\omega)$ the Laplace transform of $\gamma(t)$
(\ref{red7}). For instance, in case of ohmic friction
$I(\omega)=\bar{\gamma} m\omega$ and also for the more realistic Drude
damping (also called Debye spectral density)
$I(\omega)=m\bar{\gamma}\omega\omega_c/(\omega^2+\omega_c^2)$
one has $\gamma=\bar{\gamma}$. Since one is interested in the dynamics
on long time scales where the position distribution equilibrates, only
the low frequency modes of the residual degrees of freedom are
relevant. Now, for overdamped quantum systems given a typical
frequency $\omega_0$ of the bare
system, e.g.\ its ground state frequency, by strong damping we
mean
\begin{equation}
\frac{\gamma}{\omega_0^2}\gg \hbar\beta,
\frac{1}{\omega_c},\frac{1}{\gamma}\, .\label{smolu2}
\end{equation}
In other words, the time scale separation known from the classical
Smoluchowski limit is extended in the quantum domain to incorporate
also the time scale for quantum fluctuations
$\hbar\beta$. Correspondingly, one considers the dynamics on a coarse
grained time scale $s\gg
\hbar\beta,\frac{1}{\omega_c},\frac{1}{\gamma}$ in real time and
$\tau\gg \frac{1}{\omega_c}, \frac{1}{\gamma}$ in imaginary time. The
consequences are substantial: (i) The strong friction limit suppresses
non-diagonal elements of the reduced density matrix during the time
evolution. This simply reflects the fact that a quantum system tends
to behave more classically, the stronger coherences are destroyed in
presence of a heat bath. (ii) The real-time part of the damping kernel
$K(s)$ becomes local in time so that a time evolution equation of the
form $\dot{\rho}(t)={\cal L}\, \rho(t)$ with a time independent
operator ${\cal L}$ may exist. For the quasi-ohmic case $\omega_c\gg
\gamma$ considered in the sequel the corresponding range in parameter
space covered by
(\ref{smolu2}) is shown in fig.~2. It is well separated from the weak
friction limit region and comprises temperatures from the classical
$\gamma\hbar\beta\ll 1$ to the deep quantum domain
$\gamma\hbar\beta\gg 1$.

This way it has been shown in \cite{ankerprl} that for
 continuous systems moving in sufficiently smooth potentials $V(q)$
 the diagonal
part of the reduced density matrix $P(q,t)=\rho(q,q,t)$ obeys the equation of
motion $\dot{P}(q,t)={\cal L}_{\rm QS}\, P(q,t)$ with
\begin{equation}
{\cal L}_{\rm QS}=
\frac{1}{\gamma m} \frac{\partial}{\partial q}\left\{ V'_{\rm
  eff}(q)+\frac{\partial}{\partial q} \left[\frac{1}{\beta}+\Lambda
    V''(q)\right]\right\} \label{smolu3}
\end{equation}
where $^.$ abbreviates $d/dt$, $'$ stands for $d/dq$, and $V_{\rm
  eff}=V+\Lambda V''/2$. Here
\begin{equation}
\Lambda=\frac{2}{\hbar\beta}\, \sum_{n=1}^{\infty}\,
\frac{1}{\nu_n^2+\nu_n \hat{\gamma}(\nu_n)}\label{smolu4}
\end{equation}
with Matsubara frequencies $\nu_n=2\pi n/\hbar\beta$ measures typical
quantum fluctuations in position space. In the high temperature range
$\gamma\hbar\beta\ll 1$ these are negligible and one recovers the
classical Smoluchowski operator ${\cal L}_{\rm CLS}={\cal L}_{\rm
  QS}(\Lambda=0)$. For low
temperatures, however,
quantum effects appear and for $\gamma\hbar\beta\gg 1$ we find
$\Lambda\approx (\hbar/m\gamma) {\rm ln}(\gamma\hbar\beta/2\pi)$.
In the specific case of a
harmonic oscillator the diffusion coefficient in ${\cal L}_{\rm QS}$
becomes independent of position and differs from the classical one
just by a renormalized temperature $1/\beta\to
1/\beta+\Lambda m\omega_0^2$. Moreover, one has $\langle
q^2\rangle\approx \langle
q^2\rangle_{\rm cl}+\Lambda$.

Indeed, the strong coupling treatment even extends to
classical phase space
so that the Wigner transform $W(p,q,t)$  of the density matrix
$\rho(q,q',t)$ satisfies a quantum Fokker-Planck equation
\cite{ankerepl}. As the quantum fluctuations in position are squeezed by
friction, the quantum fluctuations in momentum are enhanced. Namely,
for $\gamma\hbar\beta\gg 1$ they grow like $\langle p^2\rangle\approx
(m\hbar\gamma/\pi) {\rm ln}(\omega_c/\gamma)$.

Based on these results we now want to derive from the exact density matrix
$\rho_{\sigma_f,\sigma_f'}(x_f,r_f,t)$ in (\ref{red2}) approximate
equations of motion for its  diagonal
elements in case of an overdamped RC. We know that in the classical realm
  $\gamma\hbar\beta\ll 1$ this leads to the so-called Zusman
equations (ZE). A
  naive expectation would be just to replace the classical operator
  for the RC motion ${\cal L}_{\rm cl}$ by its quantum version ${\cal
      L}_{\rm QS}$ meaning simply to renormalize temperature. The
  detailed calculation, however, shows that this 
    covers only part of the quantum effects and that they also affect
    the coupling
    between electronic and RC system. In particular, due to the
    coarse graining (\ref{smolu2}) the short time
    expansion  ($\gamma t\ll 1$) used in the high temperature limit
    by Garg et al.\ is not
    applicable at lower temperatures where $\gamma\hbar\beta
{\textstyle
 {\lower 2pt \hbox{$>$} \atop \raise 1pt \hbox{$ \sim$}}}
    1$. This complicates the analysis considerably.

\subsection{Equations of Motions}\label{QZE}

According to the above discussion we restrict ourselves in the sequel
to the case $x_f=0$ and consider the time evolution of
$P_{\sigma_f,\sigma_f'}(q,t)$ on the coarse grained time scale
(\ref{smol1}) starting with preparations
$\tilde{\lambda}(r_i)=\lambda(x_i=0,r_i)$.

Then, the $x_i$-integration in (\ref{rede1}) can be performed exactly
and one has for the diagonal part of the RC-reduced density
$\tilde{P}(q,t;[\chi],[\eta])=\tilde{\rho}(x_f=0,q,t;[\chi],[\eta])$
on the coarse grained time scale
\begin{equation}
\tilde{P}(q,t;[\chi],[\eta])
    =\frac{1}{\tilde{N}(t)}\,   \int dr_i\,\tilde{\lambda}(r_i)\
        \exp \left[\frac{i}{\hbar}
        \tilde{\Sigma}_{\rm ma}(q,t,r_i;[\chi],[\eta])
        \right]\label{eqm1}
\end{equation}
with the normalization
\begin{equation}
\tilde{N}(t)=\left\{\frac{2\pi \langle q^2\rangle\, [\langle
    q^2\rangle^2-S(t)^2]}{\langle
    q^2\rangle^2+S(t)^2}\right\}^{1/2}\label{norm}
\end{equation}
and the effective action
\begin{eqnarray}
\frac{i}{\hbar}\, \tilde{\Sigma}_{\rm ma}(q,t,r_i;[\chi],[\eta])&=&
-\frac{r_i^2}{2\langle q^2\rangle}-\frac{1}{2\langle
  q^2\rangle(1-\frac{S(t)^2}{\langle q^2\rangle^2})}\,
\left(q-\frac{S(t)^2}{\langle q^2\rangle} \,
r_i\right)^2\nonumber\\
&&- \Phi_r([\chi],[\eta],t)\, r_i- \Phi_q([\chi],[\eta],t)\,
q-\Phi_0([\chi],[\eta],t)\, .\label{eqm2}
\end{eqnarray}
Here, the first term describes the equilibrium distribution of the RC
with $\langle q^2\rangle\approx 1/m\beta\omega_0^2+\Lambda$  the
 variance of the overdamped RC and $S(t)$ is the
symmetrized (real) part of the position autocorrelation function
\begin{eqnarray}
\langle q(t)q\rangle&=&\frac{1}{2}\langle q(t) q+q q(t)\rangle-i
\frac{1}{2}\langle q(t) q-q q(t)\rangle\nonumber\\
&=& S(t)+i A(t)\, .\label{eqm3}
\end{eqnarray}
On the coarse grained time scale these correlation functions simplify to read
\begin{equation}
S(t)\approx \langle q^2\rangle_{\rm cl}\ {\rm e}^{-\omega_0^2
  t/\gamma}\ ,\ \ \ \ \ \ A(t)\approx -\frac{\hbar}{2 m\gamma}\ {\rm
  e}^{-\omega_0^2 t/\gamma}\, .\label{eqm4}
\end{equation}
Eventually, the time dependent coefficients $\Phi_j, j=r,q,0$ are
functionals of the electronic paths and are specified in  Appendix A.

The crucial point is now to take into account the impact  of the
sluggish RC motion onto the electronic dynamics.
For the spin held fixed,  the RC relaxes
on one of the diabatic surfaces (see also fig.~1)
\begin{equation}
V_{\eta}(q)=\frac{m\omega_0^2}{2}
\left(q-\frac{c_0}{m\omega_0^2}\ \eta\right)^2\label{eqm4a}
\end{equation}
($\eta=-\, [+]$ for donor [acceptor]), the relative position of
which is characterized by the reorganization energy
\begin{equation}
E_{\rm r}=2\, m \omega_0^2\, q_0^2\, \ \ \ \mbox{with}\ \
q_0=\frac{c_0}{m\omega_0^2}\, .\label{eqm4c}
\end{equation}
A dynamical spin causes
transitions between these surfaces that are most likely to occur in
the Landau-Zener (LZ) region around $q=0$ where the
diabatic potential surfaces intersect.
For the motion within this region the time scale of
the RC dynamics is $t_{\rm LZ}$ which is 
 the typical time a RC trajectory lives in the LZ range before
relaxing towards the donor/acceptor minima. The electronic dynamics starts
from a diagonal state, the donor,  and finally reaches the other diagonal
state, the acceptor, via jumps through the non-diagonal
states ("blips"). The bare spin dynamics can thus be visualized as
jumps along the edges of a square where the corners represent the
spin-Hilbert space. Now, this square follows the slow RC 
motion in  such a way that blips appear substantially only if
the RC lies in the LZ range, while the spin essentially rests in one
of its diagonal states away from it. For sufficiently large
reorganization energies $\beta E_{\rm r}>1$ the probability for the RC to
be activated into the LZ region is small [of the order of the
  activation rate $k_{\rm act}\approx\omega_0
  \exp(-\beta E_{\rm r})$]  and one has a separation of
time scales between the dwell times of the spin in its diagonal and
nondiagonal states, respectively, namely $t_{\rm act}=1/k_{\rm act}\gg
t_{\rm blip}$. This scenario allows for a perturbative treatment where
it suffices to assume $\hbar\beta\ll t_{\rm
  blip}\ll t_{\rm LZ}<\gamma/\omega_0^2$.

To summarize the situation: We assume an overdamped RC coupled to a
spin, activated by a
rare event into the LZ range where it diffuses for times of order
$t_{\rm LZ}$ while the spin jumps between
diagonal ($\eta=\pm 1$, $\chi=0$) and nondiagonal ($\eta=0$, $\chi=\pm
1$) states on a time scale of the order of $t_{\rm blip}$, i.e.,
\begin{equation}
\frac{1}{\gamma},\hbar\beta\ll t_{\rm blip}\ll t_{\rm LZ}<
  \frac{\gamma}{\omega_0^2}\ll t_{\rm 
  act}\, .\label{eqm5}
\end{equation}
These conditions serve as the starting point for a perturbative
approach and  will be confirmed self-consistently afterwards.

The immediate consequence of (\ref{eqm5}) is that for times $t\gg
t_{\rm blip}$ integrals in the
functionals $\Phi_j([\eta],[\chi],t)$ containing $\chi$-spin paths are
approximated by
\begin{equation}
\int_0^t ds \, \chi(s)\, f(s)\approx \int_{t-t_{\rm blip}}^{t}ds
 \,\chi(s)\, f(s) =\int_0^{t_{\rm blip}}ds \, \chi(t-s)\, f(t-s)\,
 ,\label{eqm6} 
\end{equation}
where $f(s)$ denotes smooth time dependent functions decaying to zero
on the time
scale $t_{\rm LZ}$. The typical magnitude of these integrals is thus of
order $t_{\rm blip} |f(t)|$ and therefore much smaller than integrals of
the form
\begin{equation}
\int_0^t ds \, \eta(s) f(s)\approx \eta(t) \int_0^t ds \, f(s)\label{eqm7}
\end{equation}
 which are of order $t_{\rm LZ}\, {\rm max}_{0\leq s\leq t} |f(s)|$.
Of course, the above simplifications do not apply to all possible spin
paths, but
rather only to those that are assumed to give the {\em dominant}
contributions to the path integral (\ref{red2}). Accordingly, spin dependent
terms in the action (\ref{eqm2}) of the order of $t_{\rm blip} \Lambda$
or smaller are neglected against terms of order 1 or larger (for details
see Appendix A).

Along these lines the action (\ref{eqm2}) reduces to
\begin{eqnarray}
\frac{i}{\hbar}\tilde{\Sigma}_{\rm ma}(q,t,r_i;[\chi],\eta_f)&\approx&
-\frac{r_i^2}{2\langle q^2\rangle}
-\frac{1}{2\langle
  q^2\rangle (1-\frac{S(t)^2}{\langle q^2\rangle^2})} \, \left[
  q-\frac{S(t)}{\langle q^2\rangle}\, r_i+\frac{c_0}{m\omega_0^2}\,
  \eta_f\right]^2\nonumber\\
&&+i\, q\ \frac{2 c_0}{\hbar} \, \frac{\langle q^2\rangle_{\rm cl}}{\langle
    q^2\rangle} \, \int_0^tds\ \chi(s)\, \label{eqm8}
\end{eqnarray}
where we used $\eta_f=\eta(t)$. If we take the limit $t\to 0$ with the
exact $S(t)$ [and not its coarse grained form (\ref{eqm4})]
 only the first two terms survive: Together with $\tilde{N}(t)$ the
 second one gives rise
to a $\delta$ function which  with $\eta(0)=\eta_f=-1$ restricts $r_i$
to $r_i=q+(c_0/m\omega_0^2)$ so that one regains from the first term
 the equilibrium distribution of the RC in the donor state.
Deviations from this equilibrium are described by the preparation
function $\tilde{\lambda}(r_i)$.
For finite times the first two terms describe the dynamics of an
overdamped harmonic oscillator in presence of a constant external
force $\eta_f\, c_0/m\omega_0^2$.
 The last term with the
$\chi$-electronic path can be interpreted as an effective
RC-dependent energy bias of the bare electronic system and can thus be
incorporated into an effective electronic Hamiltonian
\begin{equation}
H_{\rm EL, eff}(q)=-\frac{\hbar\Delta}{2}\,
\sigma_x-\frac{\hbar}{2}\left(\epsilon +\frac{2 c_0}{\hbar\kappa}\,
q\right)\, \sigma_z\, \label{eqm9}
\end{equation}
with $\kappa=\langle q^2\rangle/\langle q^2\rangle_{\rm cl}$.
Accordingly, in the reduced density matrix (\ref{red2}) we are now
able to re-express the path
integrals over the electronic paths in terms of
matrix elements and  obtain for its diagonal part with respect to
the RC
\begin{equation}
P_{\rm \sigma_f,\sigma_f'}(q,t)=\frac{1}{\tilde{N}(t)}\int dr_i \
  \tilde{\lambda}(r_i)\ {\rm
  e}^{\frac{i}{\hbar} \tilde{\Sigma}_{\rm
  RC}(q,t,r_i;\eta_f)}
 \ \ \langle \sigma_f|{\rm e}^{-\frac{i}{\hbar}H_{\rm EL, eff}
  t}|-\rangle \, \langle -|{\rm e}^{\frac{i}{\hbar}H_{\rm EL, eff}
  t}|\sigma_f'\rangle\label{eqm10}
\end{equation}
where $\tilde{\Sigma}_{\rm
  RC}(q,t,r_i;\eta_f)$ comprises the first two ($\chi$ independent)
  terms in (\ref{eqm8}).
By replacing $\tilde{\lambda}(r_i)|-\rangle\langle-|\to
  \sum_{\sigma_i,\sigma_i'}\,
\tilde{\lambda}_{\sigma_i,\sigma_i'}(r_i)\ |\sigma_i\rangle\langle
\sigma_i'|$  this result applies also to more general initial
electronic states where then
$\tilde{\lambda}_{\sigma_i,\sigma_i'}(r_i)$ includes also the
electronic preparation.

The final step is now straightforward. We take the time derivative
of the matrix elements $P_{\rm \sigma_f,\sigma_f'}(q,t)$ on the
left and the right hand side of (\ref{eqm10}), 
 express them on the
right hand side as derivatives with respect to $q$, and  neglect terms
acting on time scales of order $t_{\rm blip}\cdot 
(t_{\rm blip}/t_{\rm LZ})$ and shorter. This brings us to the
central result of this paper, namely, the generalization of the ZE to
include also the low temperature quantum regime, coined the generalized
Zusman equations (GZE) henceforth,
\begin{eqnarray}
\dot{P}_{--}(q,t)
    & = &   {\cal L}^-\ P_{--}(q,t)
        + \frac{i\Delta}{2}\, \left[ P_{-+}(q,t)-P_{+-}(q,t) \right]
        \nonumber\\
\dot{P}_{++ }(q,t)
    & = &   {\cal L}^+\ P_{++}(q,t)
        - \frac{i\Delta}{2}\, \left[P_{-+}(q,t)-P_{+-}(q,t)
    \right]\nonumber \\
\dot{P}_{-+}(q,t)
    & = &   {\cal L}^0 \ \, P_{-+}(q,t)
        + \frac{i\Delta}{2}\left[P_{--}(q,t)-P_{++}(q,t) \right]
        -i\, \left(\epsilon+\frac{2 c_0}{\hbar\kappa}\
          q \right)\ P_{-+}(q,t)\nonumber\\
\dot{P}_{+-}(q,t)
    & = &   {\cal L}^0 \ \, P_{+-}(q,t)
        - \frac{i\Delta}{2}\left[ P_{--}(q,t)-P_{++}(q,t) \right]
        +i\, \left(\epsilon+\frac{2c_0}{\hbar\kappa}\
          q\right) \ P_{+-}(q,t)\, .\label{eqm11}
\end{eqnarray}
Here,  the quantum Smoluchowski operators [see (\ref{smolu3})] read
\begin{equation}
{\cal L}^{\eta}=  \frac{1}{m\gamma} \frac{\partial}{\partial q}\,
        \left[ m \omega_0^2 (q
        -\eta\, q_0) + \frac{\kappa}{\beta} \,
        \frac{\partial}{\partial q}\right]\label{eqm12}
\end{equation}
 and describe the overdamped quantum dynamics of the
RC on the donor, the acceptor, and the averaged potential surfaces,
respectively. The diffusion constant is given by
$\kappa/\beta=m\omega_0^2 \langle
q^2\rangle$. Apart from the replacement of the classical ${\cal
  L}_{\rm cl}$ by its corresponding quantum operators, the effect of
quantum fluctuations also shows up in the $c_0/\kappa$ dependent coupling
terms of the non-diagonal matrix elements.
The coefficient $\kappa$ contains the equilibrium variance of a damped
harmonic oscillator
\begin{equation}
\kappa\equiv\frac{\langle q^2 \rangle}{\langle q^2
  \rangle_{cl}}=\hbar\beta\omega_0^2\, \int_{-\infty}^\infty
  \frac{d\omega}{2\pi}\ \frac{\omega\,
  \hat{\gamma}(-i\omega)}{(\omega_0^2-\omega^2)^2 +\omega^2\,
  \hat{\gamma}^2(-i\omega)}\ {\rm
  coth}(\omega\hbar\beta/2)\label{eqm12a}
\end{equation}
where $\hat{\gamma}$ denotes the Laplace transform of the classical
damping $\gamma(t)$ [see (\ref{red7})] and the integral is determined
by  the real
part of the dynamical
susceptibility. As discussed in the previous section, for  strong
damping (\ref{smol1}) this can be written as
\begin{equation}
\kappa\approx
1+m\omega_0^2\beta \, \Lambda\label{eqm12c}
\end{equation}
 with $\Lambda$ as in (\ref{smolu4}).
Exploiting the equivalence
of reaction coordinate model and spin boson model an alternative
representation is found to be
\begin{equation}
\kappa
=\frac{\hbar\beta}{2}\ \frac{\displaystyle\int^\infty_0 d\omega\
  J_{\rm
    SB}(\omega)\,\coth(\omega\hbar\beta/2)}{\displaystyle\int^\infty_0
  d\omega\ J_{\rm SB}(\omega)/\omega }\, ,\label{eqm13}
\end{equation}
where $J_{\rm SB}$ is the spectral density within the spin boson
formulation. This can be directly calculated from the spectral density
$I(\omega)$ of the RC representation \cite{garg,thoss}. For an ohmic
spectral density of the form $I(\omega)= m \gamma \omega
\exp(-\omega/\omega_c)$ with a large cut-off frequency $\omega_c\gg
\gamma$ one has in the overdamped limit $\gamma/\omega_0\gg 1$ a Drude
damping for the spin
boson system $J_{\rm SB}=(c_0^2/\gamma)\,
\omega/(\omega_0^4/\gamma^2+\omega^2)$.
 Since $\kappa>1$ for lower temperatures the effective
coupling $c_0/\kappa$ between electronic and RC system is decreased by quantum
fluctuations. Accordingly, the  distance between the
minima of the
diabatic surfaces effectively shrinks, thus reflecting the fact that
the activation barrier between donor and acceptor surface tends to
become transparent. Indeed, below in Sec.~\ref{rate} we will see that
 $\kappa$ crucially influences the transfer rate at lower
temperatures and gives rise to a renormalized reorganization energy
$E_{\rm r}/\kappa$.
We note that this is a very important point and 
 in contrast to the popular approach  where
quantum effects are taken into account {\em ad hoc} by a simple
renormalization of temperature \cite{hopfield,garg}. This latter
procedure violates the 
detailed balance condition and is apparently not consistent with the derivation
of the GZE from the exact dynamics.
Of course, in the high temperature limit
$\gamma\hbar\beta\ll 1$ the GZE (\ref{eqm12}) reduce to the ZE.

\subsection{Range of Validity}\label{valid}

To derive the above GZE we have relied on various assumptions based on
time scale separations. While for the bare RC dynamics the time scale
separation grown out of the overdamped limit is well controlled, for the
coupled electronic+RC motion the conditions (\ref{eqm5}) certainly need
to be confirmed and also be specified more precisely.

The time scale $t_{\rm LZ}$ is gained
  from the diffusion constant $D$ of the
  GZE according to $q_{\rm LZ}=\sqrt{D
  t_{\rm LZ}}$ where $l_{\rm LZ}=
  q_0\hbar\Delta/E_{\rm r}$ is the width of the Landau Zener
  region. With $D=\kappa/\gamma 
  m\beta$ we find $t_{\rm LZ}=(\gamma/\omega_0^2)
  (\Delta\hbar\beta)^2/(\beta \kappa E_{\rm r})$. To get an explicit
  expression for $t_{\rm blip}$ is more 
  involved.  For $\Delta=0$ the GZE for the nondiagonal elements
  are simply solved by calculating its corresponding
  Greens-function. This has been done for the classical case in
  \cite{cukier,goychuk} so that we
  collect in Appendix B  the final results adapted to the quantum case
  only.  The oscillating part (phase) of the Greens function
  is proportional to $c_0\, q\, t$ which is reminiscent of the
  time evolution of the nondiagonal projectors, $|+\rangle\langle
  -|$ and $|-\rangle\langle
  +|$, under the $\Delta=0$-Hamiltonian $c_0\, q\,
  \sigma_z$. Accordingly,  for
  the RC fixed (no diffusion)  in the vicinity of the
  LZ point $q=0$, the corresponding oscillation period $t_{\rm osz}$ tends to
  infinity for $\Delta=0$ and for $\Delta\neq 0$ the spin moves
  coherently through its 
  discrete Hilbert space with a typical dwell time of order $1/\Delta$
  in each state. However, when
  the RC diffuses away
  from the LZ range, $t_{\rm osz}$  decreases. In fact, when the RC
  approaches the well regions around
  $\pm q_0$,   these
  oscillations become so fast that
  on average (averaged over typical times, e.g.\ of order $t_{\rm LZ}$)
  the $P_{\pm \mp}$ are basically washed out and effectively the spin
  resides in one of its diagonal states. This way, the phase of the
  nondiagonal Greens function encodes the changeover from spin
  dynamics to spin trapping. Its time scale $t_{\rm osz}$ associated
  with  the diffusive RC motion,
  i.e.\ $|q|\approx \sqrt{D t}$, is obtained as
\begin{equation}
t_{\rm osz}\approx \left(\frac{\gamma}{\omega_0^2}\,
\frac{\hbar^2\beta^2}{\beta E_{\rm r}}\right)^{1/3}\, .\label{val1}
\end{equation}
The typical dwell time in a
  nondiagonal state  may consequently be considered as $t_{\rm blip}\approx
  t_{\rm osz}$.
Obviously, for very large $\gamma/\omega_0$ when the RC dynamics
almost comes to rest (basically no diffusion), $t_{\rm osz}$ may even exceed
$1/\Delta$ and the situation discussed above is regained; the
  coherent spin dynamics lasts over large
  periods of times
  and the typical dwell time is given by $t_{\rm blip}\approx
  1/\Delta$. In this case quantum fluctuations in the RC are
  negligible and the ZE dynamics is valid anyway.
Hence, in the sequel we estimate $t_{\rm blip}=t_{\rm osz}$ with $t_{\rm
  blip}\ll t_{\rm LZ}$.

Let us now come back to our assumptions for the perturbative analysis.
The time coarse graining
(\ref{smolu3}) together with the conditions (\ref{eqm5}) lead first to the
prerequisite
for overdamped motion $\gamma/\omega_0\gg 1$ and second to 
\begin{equation}
 \frac{\hbar^2
  \beta^2}{\gamma/\omega_0^2}\ll \beta E_{\rm r}\ll
\frac{\gamma}{\omega_0},\, \frac{\Delta}{\omega_0^2/\gamma}\, 
  \Delta\hbar\beta\, .\label{val2} 
\end{equation}
By rearranging the terms and taking also into account $\beta E{\rm
  r}>1$ we arrive at the more convenient forms
\begin{eqnarray}
 \frac{E_{\rm r}}{\hbar\Delta}\
\frac{1}{\Delta\hbar\beta}\, ,\, \left(\frac{\hbar\Delta}{E_{\rm r}}\
\Delta\hbar\beta\right)^{1/2} 
&\ll &
\frac{\Delta}{\omega_0^2/\gamma}\, ,\nonumber\\
 1 < \beta E_{\rm r}&\ll &
\frac{\gamma/\omega_0^2}{\hbar\beta} \, . \label{val3}
\end{eqnarray}
Within the overdamped domain $\gamma/\omega_0\gg 1$  these relations
define the range of validity for the GZE. 
As we show in Appendix A it is exactly this range where the
simplifications of the effective action described above can be
applied in the sense of a semiclassical perturbation theory.
In particular, the first of the above conditions determines a lower 
bound on temperature depending on the ratio $E_{\rm r}/\hbar\Delta$.
Physically, with decreasing $T$
 tunneling of the RC becomes important that is incorporated in
the GZE only in a kind of static approximation (see Sec.~\ref{rate}) and
the range of validity shrinks accordingly, see fig.~3.
In comparison the ZE have additionally to obey $\gamma\hbar\beta\ll 1$
and consequently fail for much higher temperatures.
The first condition is
also readily translated into the spin boson model: for a Debye spectral
density (Drude model) the full bath cut-off frequency $\Omega_c$ is
identified as $\Omega_c\approx \omega_0^2/\gamma$ and the
spin/full-bath coupling $\alpha$ as $\alpha=E_{\rm r}/2$. Then, we
observe that the 
electronic dynamics of the GZE extends from the adiabatic
$\Delta/\Omega_c\gg 1$ to the nonadiabatic limit $\Delta/\Omega_c\ll 1$
also at lower temperatures (see fig.~3).
 The second relation is independent of $\Delta$ and gives
an upper bound for the reorganization energy. This is understood by
recalling that  a larger $E_{\rm r}$ leads to steeper potential surfaces
at the LZ point causing larger momentum transfers.  In order for the
Smoluchowski reduction to be valid, the damping must increase to
compensate for that. The same  restriction holds in the classical regime
and may thus explain recent discussions about negative probabilities
etc.\ in the ZE for higher reorganization energies
\cite{frantsuzov}. In particular,  
it reveals that in a strict sense the GZE/ZE cannot be used for ET in
polar solvents where  reorganization energies are relatively large \cite{rem}
(for a detailed discussion of this point see \cite{goychuk3}).

Qualitatively, the electronic dynamics described by the GZE is found to be
either coherent or incoherent. The first case applies always for
sufficiently large $\Delta/(\omega_0^2/\gamma)\gg 1$ when the RC motion
tends to be trapped by friction so that effectively a static field,
distributed according to its equilibrium distribution, is coupled to the
spin. As known from the classical limit (ZE) this evokes damped coherent
oscillations in the electronic occupation probabilities
\cite{lucke97}. For smaller ratio 
$\Delta/(\omega_0^2/\gamma)$ and sufficiently $\beta E_{\rm
  r}>1$ one reaches the domain of incoherent decay characterized by
transfer rates. To derive these from the GZE will be the subject of the
last section.

\section{Equations of Motion in Phase Space}\label{fokker}

As we have seen above, for higher reorganization energies the time
coarse graining of the RC motion runs into conflict with the electronic
dynamics. The path integral formulation allows to extract a
set of equations in full phase space by keeping also nondiagonal
elements of the reduced density matrix with respect to the RC
($x_f\neq 0$). Since the RC is just a damped harmonic oscillator
(coupled to dynamical electronic paths) everything goes through
exactly. As already mentioned above, for continuous systems this
generalization of the quantum Smoluchowski equation to a quantum
Fokker-Planck equation has already been done in \cite{ankerepl}. In
particular,  one can thus explicitly take into account the
equilibration process of the momentum.

Here, starting again with (\ref{rede1}) and working along the lines
described above, we find for the $\chi$-independent part of  the action
\begin{eqnarray}
\frac{i}{\hbar}\tilde{\Sigma}_{\rm RC}(x_f,r_f,r_i;\eta_f,t)&=&
  -\frac{r_i^2}{2\langle
  q^2\rangle} -i \frac{m\omega_0^2}{\hbar\gamma}\, x_f
  r_f-\frac{\langle p^2\rangle}{2\hbar^2}\, x_f^2\nonumber\\
&&-\frac{1}{2\langle q^2\rangle \left(1-\frac{S(t)^2}{\langle
      q^2\rangle}\right) }\, \left[r_f-\frac{S(t)}{\langle q^2 \rangle} r_i
        -i\frac{2\omega_0^2 \langle q^2 \rangle}{\hbar\gamma} x_f
        +\frac{c_0}{m\omega_0^2}\, \eta_f\right]^2\, .\label{fp1}
\end{eqnarray}
In the low temperature range $\langle p^2\rangle \approx
(m\hbar\gamma/\pi){\rm ln}(\omega_c/\gamma)$ so that, as expected,
nondiagonal elements in the RC are strongly suppressed and $|x_f|$
takes only small values, roughly of order $1/\sqrt{\gamma}$. The
$\chi$-dependent parts 
of the action
are approximated as above with the
result that to this order of perturbation theory (neglecting terms of
order $t_{\rm blip}\Lambda$ or smaller) no additional $x_f\, \chi$
dependent terms have to be retained.

In a next step we go over to a phase space formulation by taking the
Wigner transform of $\rho_{\sigma_f,\sigma_f'}(x_f,r_f,t)$, i.e.,
\begin{equation}
W_{\sigma_f,\sigma_f'}(p,q,t)=\frac{1}{2\pi\hbar}\int dx_f \,
\rho_{\sigma_f,\sigma_f'}(x_f,q,t)\ {\rm e}^{-\frac{i}{\hbar} p
  x_f}\, .\label{fp2}
\end{equation}
The corresponding equations of motion for the matrix elements of  the
Wigner density matrix are then found as
\begin{eqnarray}
\dot{W}_{--}(p,q,t)
    & = &   {\cal L}^-_{\rm QFP}\ W_{--}(p,q,t)
        + \frac{i\Delta}{2}\, \left[
    W_{-+}(p,q,t)-W_{+-}(p,q,t) \right]
        \nonumber\\
\dot{W}_{-+}(p,q,t)
    & = &   {\cal L}^0_{\rm QFP} \ \, W_{-+}(p,q,t)
        + \frac{i\Delta}{2}\left[W_{--}(p,q,t)-W_{++}(p,q,t) \right]\nonumber\\
    &&\hspace{5cm}    -i\, \left(\epsilon+\frac{2c_0}{\hbar\kappa}\
          q\right) \ W_{-+}(p,q,t)\, \label{qfp}
\end{eqnarray}
where the equation for $W_{++}$ follows from the first one by
replacing $+$ with $-$ and vice versa, while that for  $W_{+-}$
follows from the second one by  interchanging $+$ and $-$ and complex
conjugation of its last term.

Compared to the GZE, in the above equations the quantum Smoluchowski
operators are replaced by quantum Fokker-Planck operators
\begin{equation}
{\cal L}^{\eta}_{\rm QFP}=  \frac{\partial}{\partial p}\,
        \left[ m \omega_0^2 (q
        -\eta\, q_0) + \gamma
        p\right]-\frac{p}{m}\frac{\partial}{\partial p}
        +\gamma \langle p^2\rangle \frac{\partial^2}{\partial
        p^2}
+\left[\langle q^2\rangle-\frac{\langle p^2\rangle}{m}\right]\,
        \frac{\partial^2}{\partial p\partial q}\, .
\label{fpop}
\end{equation}
They differ from classical Fokker-Planck operators \cite{risken} mainly by
the last term that describes coupled $pq$ diffusion and has also been found
for  weak damping \cite{haake,karrlein}. In the high temperature
range $\gamma\hbar\beta\ll 1$  its
diffusion constant tends to zero due to the equipartition theorem and
one recovers from (\ref{qfp}), (\ref{fpop}) the phase space equations
derived by
Garg et al.\  \cite{garg}. For low temperatures
 $\gamma\hbar\beta\gg 1$ its impact becomes substantial and is
dominated by the growing $\langle p^2\rangle$ variance. Numerical results
with the quantum Fokker-Planck equation for a simple harmonic
oscillator \cite{ankerepl} 
show that for strong damping mean values like $\langle p(t)\rangle$ and
correlation functions like $\langle p(t)p\rangle$ are in excellent
agreement with exact results and that small deviations only occur in the
short time range. Of course, if the momentum is eliminated adiabatically
from ({\ref{fpop}) we regain the GZE.

With the exact equilibrium variances inserted into (\ref{fpop}) the
corresponding Fokker-Planck operator is even more
general \cite{karrlein}: It
gives for the set of
initial conditions (\ref{et6}) and ohmic friction (supplemented by a high
frequency cut-off) the exact quantum dynamics of a damped harmonic
oscillator in the long time limit independent of temperature and
damping strength. While in principle the range of validity of
(\ref{qfp}) is defined by  the conditions (\ref{val3}),
one might thus be tempted to think that with the ``exact''
 ${\cal L}_{\rm QFP}$ one could also extend this range. However, a
deeper analysis reveals that by weakening for lower temperatures one
of the conditions in 
(\ref{val3}) 
 additional $\chi$ and $\eta$
dependent terms in the effective action  [see (\ref{eqm2})] must be
taken into account.  Equations of motion with time independent
coefficients are then out of reach.
  This suggests that the relations (\ref{val3}) define
necessary and sufficient criteria for the existence of (simple)
dynamical equations for the coupled electronic/RC motion in the strong
damping domain at lower temperatures. Work to study the quantum phase
space dynamics in 
detail is in progress and will be presented elsewhere.

\section{Quantum Transfer Rates}\label{rate}

Electron transfer rates have been calculated analytically and
numerically within the RC and the spin boson representation as well.
In essence, two limiting cases are distinguishable, in particular since they
allow for perturbative analytical treatments (cf.~fig.~3). Roughly
speaking, in the
nonadiabatic regime the RC crosses the LZ region fast compared to the
dynamics of the spin ($\Delta/\Omega_c\ll 1$, $\Omega_c $ cut-off
frequency of the full bath) so that golden rule
techniques in the
electronic coupling $\Delta$ can be invoked. This way,
transfer rates 
for classical and quantum mechanical baths have been obtained
\cite{wolynes,coalson,weiss}. In the 
opposite case of a very slow RC compared to the bare electronic motion
 ($\Delta/\Omega_c\gg 1$) the sluggish RC dynamics controls the
transfer and established methods like e.g.\ Kramers flux over population
approach or transitions state theory on the lower {\em adiabatic} surface
are applicable \cite{marcus,goychuk}. A crucial point is the
changeover between purely 
adiabatic to nonadiabatic dominated transfer. As first shown by Zusman
\cite{zusman} and then later by Garg et al.\ \cite{garg} the ZE allow to
derive an explicit expression for an interpolation
  formula that reduces to the adiabatic (nonadiabatic) rate constant
  in the respective limits. A detailed survey over  the available analytical
  results and their performance compared to real-time Quantum Monte Carlo data
  was given recently in \cite{lothar}: For $\Delta/\Omega_c\ll 1$
  the nonadiabatic rate expression works quite well over the whole
  temperature range. The same is true for $\Delta/\Omega_c{\textstyle
 {\lower 2pt \hbox{$>$} \atop \raise 1pt \hbox{$ \sim$}}} 1$ and
higher temperatures $\Delta\hbar\beta{\textstyle
 {\lower 2pt \hbox{$<$} \atop \raise 1pt \hbox{$ \sim$}}} 1$ where the
Zusman rate is
valid. The only domain where no explicit rate formula is known so far
is $\Delta/\Omega_c{\textstyle
 {\lower 2pt \hbox{$>$} \atop \raise 1pt \hbox{$ \sim$}}} 1$  at lower
temperatures
$\Delta\hbar\beta {\textstyle
 {\lower 2pt \hbox{$>$} \atop \raise 1pt \hbox{$ \sim$}}} 1$.
It is exactly this region where the GZE apply.

The procedure to extract the rate from the ZE has been outlined in
\cite{cukier,goychuk}. It also works for the GZE so that details
are omitted here. By solving formally the two equations for the
nondiagonal elements (using the proper Greens function, see Appendix B)
and plugging this result into the two equations for the diagonal ones, we
arrive at two coupled integro-differential equations. These can be
reduced to two coupled partial differential equations by exploiting the
time scale separation $t_{\rm blip}\ll t_{\rm LZ}$ to end up with
\begin{eqnarray}
\dot{P}_{--}(q,t)
    &=& -K(q)\left[P_{--}(q,t) -P_{++}(q,t)\right]
        +{\cal L}^- P_{--}(q,t)\nonumber \\
\dot{P}_{++}(q,t)
    &=& K(q)\left[P_{++}(q,t) -P_{--}(q,t)\right]
        +{\cal L}^+ P_{++}(q,t)\, .\label{rate0}
\end{eqnarray}
Here, the effective position dependent electronic coupling turns out
to be
\begin{equation}
K(q)=\frac{\Delta^2}{2}\, {\rm Re}\,
\int_{-\infty}^{\infty}dq'\int_0^\infty dt\ G_0(q',t|q)\label{rate0b}
\end{equation}
where $ G_0(q',t|q)$ is the Greens function for the diffusion in the
average potential that determines the nondiagonal elements.
Next, the equations (\ref{rate0}) are Laplace transformed (with
respect to time) and projection operator techniques provide a mapping
onto electronic populations. Finally, the transfer rate follows in the
long time limit, i.e.\ for Laplace parameter tending to zero, in the form
\begin{equation}
k_+=\frac{k_+^{\rm na}}{1+k_+^{\rm na}/k_+^{\rm d}+k_-^{\rm na}/k_-^{\rm
d}}\, \label{rate1}
\end{equation}
for the  forward rate where $k_\pm^{\rm na}$ and $k_\pm^{\rm d}$
are so-called nonadiabatic and
diffusive rate constants for forward and backward transfer, respectively.
These are related by the  detailed balance condition
\begin{equation}
k_+^\alpha/k_-^\alpha={\rm e}^{\beta\hbar\epsilon}\label{rate2}
\end{equation}
with $\alpha$=na, d. The backward rate $k_-$ follows from $k_+$
by interchanging $+\leftrightarrow -$.
Specifically, one obtains
\begin{equation}
k_\pm^{\rm na}=\frac{1}{2\pi\langle q^2\rangle}\,
\int_{-\infty}^\infty dq \ K(q)\,  \exp\left[-\frac{(q\pm q_0)^2}{2\langle
    q^2\rangle}\right] \label{rate0c}
\end{equation}
which represents the nonadiabatic rate as an average of the effective
electronic coupling over the equilibrium distribution of the donor and
acceptor surfaces, respectively. The diffusion in these harmonic wells
is described by the rate constants
\begin{equation}
\frac{1}{k_\pm^{\rm d}}=\int_0^\infty dt\ \left\{G_\mp(q_0,t|q_0)\,
  \sqrt{2\pi\langle q^2\rangle}\, \exp\left[\frac{(q\pm q_0)^2}{2\langle
    q^2\rangle}\right]-1\right\} \label{rate0d}
\end{equation}
where $G_\mp(q_0,t|q_0)$ denote the Greens functions for the RC motion in
the  donor and acceptor (see Appendix B).
Both expressions can be simplified further for a sufficiently large
reorganization energy $\beta E_{\rm r}>1$.
It then turns out that the nonadiabatic rates coincide  with
 golden rule rates, i.e.,
\begin{equation}
k_\pm^{\rm na}=\frac{\hbar\Delta^2}{4}\ \sqrt{\frac{\pi
    \beta}{E_{\rm r}/\kappa}}\
\exp\left(-\beta E_\pm^\# \right)\, ,\label{rate2b}
\end{equation}
where
\begin{equation}
E_\pm^\#=\frac{(E_{\rm r}/\kappa\mp\hbar\epsilon)^2}{4
E_{\rm r}/\kappa}\label{rate2c}
\end{equation}
are the effective activation energies for the forward/backward rates.
The diffusive rates are identical to the overdamped Kramers
rates for escape on the lower adiabatic surface
\begin{equation}
k_\pm^{\rm d}= \frac{\omega_0^2}{\gamma}\, \sqrt{\frac{\beta E_\pm^\#}{ 4
\pi\kappa}}\ \exp\left(-\beta E_\pm^\#\right)\, .\label{rate3}
\end{equation}
Accordingly, for the
important case of a symmetric transfer ($\epsilon=0$) the ultimate
rate for equilibration, i.e.\ the
total  rate $k=k_++k_-$, takes the
handsome form
\begin{equation}
k=\frac{\Delta^2}{1+ g}\ \sqrt{\frac{\hbar^2 \pi \beta}{4\,E_{\rm
      r}/\kappa}}\
\exp\left(-\beta E_{\rm r}/4\kappa\right)\,
\label{rate4}
\end{equation}
with an adiabaticity parameter
\begin{equation}
g=\pi\,\kappa\ \frac{\Delta}{\omega_0^2/\gamma}\
\frac{\hbar\Delta}{E_{\rm r}}\, .\label{rate5}
\end{equation}
For $g\gg 1 $ one recovers the adiabatic, for $g\ll 1 $ the
nonadiabatic rate constant [cf.~(\ref{val3}) and fig.~3].
The above rate expression looks like the classical Marcus/Zusman result with a
renormalized reorganization energy $E_{\rm r}\to E_{\rm
  r}/\kappa$. Note that this
simple renormalization only appears if quantum fluctuations in the
$c_0$-coupling terms of the equations for $P_{\pm\mp}$ [see (\ref{eqm11})] are
properly taken into account. Since $\kappa\geq 1$ quantum fluctuations
always {\em reduce} the effective energy barrier which is to be surmounted.
Of course, in the high temperature limit ($\kappa\to 1$) we regain the
known result,
for lower temperature, however,  significant deviations are observed.
 (i) The ratio $\kappa$ grows with decreasing temperature, at very low
temperatures roughly linearly with $\beta$. As a consequence, keeping all
other parameters fixed, $g$ becomes larger with lower $T$ meaning that
one approaches the range where the transfer is dominated by adiabatic
processes already for smaller values of $\Delta$. This behavior of $g$ is
also consistent with the larger range of validity of the GZE at lower $T$
compared to the classical case. Moreover, it agrees with recent numerical
indications \cite{stockburger}. (ii) At lower $T$ the exponent in the
rate expression $\beta E_{\rm r}/\kappa$ tends to become temperature
independent in contrast to the classical result. In particular, for an
ohmic secondary bath $I(\omega)=m\gamma\omega\exp(-\omega/\omega_c)$ with
large cut-off frequency we gain from (\ref{eqm12a}) for large friction
$\gamma/\omega_0\gg 1$
\begin{equation}
\kappa\approx 1+\frac{\hbar\beta\omega_0^2}{\pi\gamma}\,
\left[\psi\left(1+\hbar\beta\gamma/2\pi\right)-
  \psi\left(1+\hbar\beta\omega_0^2/2\pi\gamma\right)\right]
\label{rate6}
\end{equation}
where $\psi$ denotes the logarithmic derivative of the $\Gamma$-function.
The corresponding quantum enhancement $\kappa/\beta$ is depicted in
fig.~4 for various values of the damping strength. Since $\kappa$ enters
the exponent in (\ref{rate4}), even relatively small deviations from the
classical behavior $\kappa/\beta=1/\beta$ substantially influence the rate.
 In fact, if  formally the limit of very
low temperatures $\gamma\hbar\beta\gg 1, \omega_0\hbar\beta\gg
\gamma/\omega_0$ is
taken [which in a strict sense
   is out of the range of validity
 (\ref{val3})], $\kappa/\beta$ saturates and
the exponent in (\ref{rate4}) becomes
\begin{equation}
\frac{\beta E_{\rm r}}{4\kappa} \to  \frac{\pi\gamma}{8\omega_0\, {\rm
  ln}(\gamma/\omega_0)}\,
  \frac{ E_{\rm
  r}}{\hbar\omega_0}\label{kap}
\end{equation}
which is identical to
  the overlap of two harmonic ground state wave packets
 with overdamped variance $(2\hbar/m\gamma)\,{\rm ln}(\gamma/\omega_0)/\pi$
 localized around $\mp q_0$ in the donor and acceptor wells, respectively.
Hence, nuclear tunneling is included in the above rate formula, at
least in a nondynamical way. In comparison with precise Monte Carlo
  results (from \cite{lothar}) a remarkable agreement is seen
  over a broad temperature range 
  from low up to high temperatures (fig~5). The effect of nuclear
  tunneling is clearly observable at low $T$ where the classical rates
  expression is far off. 
Only at very low $T$ and
  for $\Delta/(\omega_0^2/\gamma){\textstyle
 {\lower 2pt \hbox{$<$} \atop \raise 1pt \hbox{$ \sim$}}} 1$, i.e.\
  close to the boundary of the 
  range of validity (cf.~fig.~3),  does
 the quantum Zusman rate overestimate the true rate
  (fig.~6) but is still much better than the classical rate. In this
  parameter range
  the dynamics of the nuclear tunneling process  
 is strongly
affected by dissipation which is known to reduce the quantum rate
substantially. To complete this discussion, we note that according to
  the numerical results in \cite{lothar} for very low temperatures
  $\Delta\hbar\beta\gg 1$ and large $\Delta/(\omega_0^2/\gamma)\gg 1$
  (cf.~fig.~3) transient relaxation dynamics persists up to very long
  times and the equilibration cannot be described by
  single-exponential decay. Hence, combining the golden rule expression
  \cite{weiss,coalson}  with the above quantum Zusman rate 
  covers almost the entire parameter range where a relaxation rate
  can be found at all.

\section{Conclusions}\label{conclu}

In this work we have studied electron transfer between a donor and acceptor
site in strongly condensed phase and at lower temperatures where quantum
effects are essential. Based on an exact path integral representation
of the reduced density matrix of the electron/reaction coordinate
compound we have employed recent results on the overdamped limit in the
quantum regime to  derive equations of motions for its diagonal part
in the RC, named GZE. Utilizing the Wigner transform this was
extended to include also its nondiagonal part and to obtain a description
in phase 
space. Accordingly, we have given a rigorous derivation of
time evolution equations for an overdamped collective vibronic degree of
freedom  in presence of electronic transitions over a wide parameter
range. As another important result transfer rates have been gained from
the GZE that interpolate between adiabatic and nonadiabatic dominated
relaxation also in the low temperature domain.

The fundamental feature that makes the exact non-Markovian quantum
dynamics effectively to behave Markovian, is a time scale separation in
the overdamped RC dynamics influencing the electronic motion as well.
The main findings are the following: (i) The GZE reduce to the ZE for
high temperatures. (ii) At lower temperatures 
quantum fluctuations lead to a renormalized temperature {\em and} a
renormalized RC-spin coupling that combines to a renormalized
reorganiztion energy and thus conserves the detailed balance
condition. (iii) The range of validity of the GZE can 
be specified in detail and comprises adiabatic and nonadiabatic domains
also for $\Delta\hbar\beta>1$. In particular, the Smoluchowski reduction
requires reorganization energies bounded from above by the damping
strength. (iv) The phase space Fokker Planck equations include explicitly
the momentum equilibration of the RC, but it seems to us that they do not
allow for a substantial extension of the range of validity beyond that of
the GZE. (v) A generalization of the Marcus/Zusman expression for the electron
transfer rate contains a renormalized reorganization energy such that in
the low temperature range the rate tends to become temperature
independent. It thus describes (at least partially) nuclear
tunneling in agreement 
with precise numerical date. The golden rule formula
(see e.g.\ \cite{weiss,coalson}) and the quantum Zusman result
 give now a sufficiently precise description of the incoherent
 transfer throughout most of the relevant range of parameters.

What we have not discussed here in detail is the dynamics of the GZE and
the quantum phase space equations based on numerical solutions. This is
certainly an interesting subject, particularly for the latter one, and is
presently under study. As discussed above, the quantum
Smoluchowski limit 
has also been analyzed for anharmonic potential fields so that GZE for
these cases may also be found \cite{goychuk2}. In any case, the appealing
feature of 
equations of motions is the straightforward procedure to solve them.
Typically they are not plagued seriously by numerical instabilities and
can therefore be used to describe the ET dynamics also for very long
times. In contrast, while e.g.\ real time quantum Monte Carlo methods
provide a numerically exact scheme,  quantum noise in form of the
so-called dynamical sign problem sets a severe limit on the simulation
time. Further, the GZE give in some aspects a deeper understanding of the
underlying physics since they allow for approximate analytical results.

\section{Acknowledgements}

The authors thank L.\ M\"uhlbacher, I.\ Goychuk, H. Grabert, and E.\ Pollak for
helpful discussions. Financial support from the DFG (Bonn) through SFB276
and  AN336/1 is gratefully acknowledged.

\section*{Appendix A: Effective Action}

The exact action coefficients $\Phi_j$ in (\ref{eqm2}) are read off
from the results 
of \cite{lucke}. On the coarse grained time scale
$\hbar\beta, 1/\gamma\ll \gamma/\omega_0^2$ (Smoluchowski limit) they
are given by the following approximate expressions
\begin{eqnarray}
\Phi_r
&\approx&
\frac{2 q_0 }{\langle q^2 \rangle }\frac{s(t)}{\hbar \beta}\ \left\{
\int_0^t du \, \eta(u) A(2t-u)
+i\,
\frac{2\gamma/\omega_0^2}{\langle q^2 \rangle_{cl}\, (\hbar \beta)^2}
\int_0^tdu\, \chi(u) \left[
A(2t-u) - \kappa A(u) \right]\right\}\nonumber \\
\Phi_q
&\approx& i\,
\frac{4q_0}{ \langle q^2 \rangle_{cl}}
\frac{\gamma/\omega_0^2}{(\hbar \beta)^2}
\int_0^t du \, \chi(u) \left\{
s(t) \frac{A(t-u)}{\langle q^2 \rangle}
+\left[ 1- s(t)\right] \frac{\hbar\beta}{\gamma/\omega_0^2}\frac{A(u)}{A(t)}
\right\}
\nonumber \\
&&
+\frac{2q_0}{\langle q^2 \rangle_{cl} \langle q^2 \rangle}
\frac{s(t)}{\hbar \beta}
\int_0^t du\, \eta(u) A(t-u)
\nonumber\\
\Phi_0
&\approx&i\,
\frac{4 q_0^2}{\langle q^2 \rangle_{cl}}\frac{s(t)}{\hbar \beta
  (\gamma/\omega_0^2)} \, 
\int_0^t du\,\int_0^t ds\, \chi(u)\eta(s)
\left[ \frac{ A(u)}{A(s)} - \frac{ A(2t)}{A(u+s)}
-\theta(s-u) \frac{A(u)}{s(t) A(s)} \right]
\nonumber \\
&&
-\frac{q_0^2}{\langle q^2 \rangle_{cl} \langle q^2 \rangle}
\frac{s(t)}{(\gamma/\omega_0^2) \hbar \beta}
\int_0^t \int_0^t ds\, du\, \eta(s) \eta(u)A(2t-u-s)
\nonumber \\
&&
+\frac{2q_0^2}{\langle q^2 \rangle_{cl} \langle q^2 \rangle}
\frac{1}{(\hbar\beta)^2}
\int_0^t du \int_0^t ds\,  \chi(u) \chi(s)
\nonumber \\
&& \times \left\{ \kappa^2 \left[ s(t)-1 \right] \frac{S(s)
  S(u)}{S(2t)} -2 
\kappa \left[ s(t)-\theta(s-u) \right] S(s-u) + s(t) S(2t-s-u)
\right\}\label{coeff1}
\end{eqnarray}
with the abbreviation $s(t)=\left(1-\frac{S^2(t)}{\langle q^2
\rangle^2}\right)^{-1}$ and $A(t)$ and $S(t)$ as in (\ref{eqm4}).
If also the  time scale separation $ \hbar \beta\ll t_{\rm blip}\ll
t_{LZ}\ll t_{\rm act}$  is applied, these coefficients reduce further. In
a first step all leading order terms in $t_{\rm
  blip}$ are kept 
\begin{eqnarray}
\Phi_r
&\approx&
\frac{2 q_0}{\langle q^2 \rangle^2 }\frac{\gamma/\omega_0^2\,
  }{\hbar \beta} \, A(t)\, s(t)\, \left\{ \left[1-A(2t)/A(t)\right]\eta_f
-i
 \frac{2\kappa\, \Lambda}{\langle q^2 \rangle_{cl}\hbar\beta}
 \int_0^t ds\,\chi(s) \right\}\nonumber\\
\Phi_q &\approx& \frac{2 q_0}{\langle q^2 \rangle}\, \left[s(t)\,
\eta_f+i\frac{\langle
    q^2\rangle - \Lambda s(t)}{\langle
    q^2\rangle_{cl}\hbar \beta}\, \int_0^t ds\,  \chi(s)\right]\nonumber\\
\Phi_0
&\approx&i\,
\frac{4q_0^2}{\langle q^2
  \rangle^2_{cl}}\frac{\gamma/\omega_0^2}{(\hbar \beta)^2}
 \left\{ \frac{\Lambda}{\langle q^2 \rangle} s(t)
\left[ A(0) - A(t) \right] -\frac{t_{\rm blip}}{2}
\frac{A(0)}{\gamma/\omega_0^2} \right\}
\eta_f\,
\int_0^t ds\, \chi(s) \\
&& -\frac{q_0^2}{\langle q^2 \rangle_{cl} \langle q^2 \rangle}
\frac{\gamma/\omega_0^2}{\hbar \beta} \, s(t)\, \left[ A(0) - 2A(t) +
A(2t)\right] \eta_f^2\nonumber\\
&&+ \frac{2 q_0^2}{\langle q^2\rangle_{cl} \langle q^2\rangle}
\frac{\Lambda}{\hbar\beta}\frac{\langle
    q^2\rangle - \Lambda s(t)}{\langle
    q^2\rangle_{cl}\hbar \beta}\, \int_0^tds\int_0^tdu \, \chi(s)\,
    \chi(u)\label{coeff2}
\end{eqnarray}
with $A(0)=-\hbar/2m\gamma$ from (\ref{eqm4}).
To extract in a second step the relevant terms, we estimate the order of
magnitudes of the various contributions. There, $\Phi_r$ and $\Phi_q$
carry an overall factor $\beta E_{\rm r}/q_0$, $\Phi_0$ a factor $\beta
E_{\rm r}$. Specifically, one obtains for $\Phi_r$: The 
$\eta_f$ term is of order $\beta E_{\rm r}/q_0$, the $\chi$ term of order
$\xi\, \beta E_{\rm r}/q_0$ with
\begin{equation}
\xi= \frac{\omega_0 t_{\rm blip}}{\kappa}
\,\frac{\Lambda}{\hbar/m\omega_0}\, .
\end{equation}
Next $\Phi_q$: The $\eta_f$ term is of order $\beta E_{\rm r}/q_0$; the
first $\chi$ term is of order $t_{\rm blip}/(\hbar\beta\kappa)\, \beta
E_{\rm r}/q_0$, the second one of order $\xi\, \beta E_{\rm r}/q_0$.
Finally, the $\Phi_0$ contribution: In the first line terms are of order
$\beta E_{\rm r} \xi$ or smaller, the second line is of order $\beta
E_{\rm r}$, and the third one of order $\beta E_{\rm r} \xi \, t_{\rm
blip}/\hbar\beta$. It is now assumed that  $\beta E_{\rm r} \xi \, t_{\rm
blip}/\hbar\beta\ll 1$ so that all terms of this and smaller order (e.g.\
$\beta E_{\rm r} \xi $) are negligible in the sense of a semiclassical
approximation. Accordingly, the expression (\ref{eqm8}) for the effective
action follows. In fact, within the range of validity of the GZE
specified in (\ref{val3}) this assumption is indeed justified, thus
proving the self-consistency of the approach.

\section*{Appendix B: Greens functions}

The Greens function for the averaged potential is obtained
from solving
\begin{equation}
\left[\frac{\partial}{\partial t} -{\cal L}^0
+ i\left( \epsilon + \frac{2 c_0}{\hbar \kappa} q \right) \right]
 G_0(q,t-t'|q')
    = 0
\end{equation}
with the initial condition
$\lim\nolimits_{t \to 0} G_0 (q,0|q')=\delta (q-q')$
and the boundary conditions $G(\pm \infty,t-t'|q')=0$.
Details of the calculation have been reported in
\cite{cukier,goychuk}. Here, we 
specify the final result
\begin{equation}
G_0(q,t|q')
    =   \frac{1}{2 \sqrt{\pi a(t)}}\exp \left\{
        -\frac{\left[q-ib(q',t)\right]^2}{4 a(t)} +c(q',t)
    \right\}\, , \label{appa1}
\end{equation}
where
\begin{eqnarray}
a(t)
    &=& \frac{\langle q^2 \rangle}{2}\left(
        1-{\rm e}^{-2\frac{\omega_0^2}{\gamma}t}\right)\nonumber\\
b(q',t)
    &=& -\frac{\omega_0^2}{\gamma}\langle q^2 \rangle
        \frac{2}{\hbar}\frac{c_0}{\kappa}
        \left(1-{\rm e}^{-\frac{\omega_0^2}{\gamma}t}\right)^2
        +i q' {\rm e}^{-\frac{\omega_0^2}{\gamma}t}\nonumber \\
c(q',t)
    &=& \frac{\gamma^2}{\omega_0^4}\langle q^2 \rangle
        \frac{2}{\hbar^2}\frac{c_0^2}{\kappa^2}
        \left({\rm e}^{-2\frac{\omega_0^2}{\gamma}t}
        -4{\rm e}^{-\frac{\omega_0^2}{\gamma}t}
        +3-2\frac{\omega_0^2}{\gamma}t\right)
        +i\frac{\gamma}{\omega_0^2}
        \left(1-{\rm e}^{-\frac{\omega_0^2}{\gamma}t}\right)
        \frac{2 c_0}{\kappa\hbar}q'\, .\label{appa2}
\end{eqnarray}
To estimate $t_{\rm blip}$ (see Sec.~\ref{valid}) we consider times $t\ll
t_{\rm LZ}=\gamma/\omega_0^2$ and retain the imaginary part of the
exponent  in leading order given by the imaginary part to $c(q',t)$.

The Greens functions for the donor/acceptor surfaces are gained from
 $\left[ \partial_t - {\cal L}^\mp \right]G^\mp =0$ and read
\begin{equation}
G_{\mp}(q,t-t'|q')
=   \frac{1}{2 \sqrt{\pi a(t)}}\exp \left\{
  -\frac{\left[q\pm q_0
    -(q\pm q_0){\rm e}^{-\frac{\omega_0^2}{\gamma}(t-t')}
    \right]^2}{4a(t)}\right\}\label{appa3}
\end{equation}

{}

\pagebreak

\begin{center}
{\large FIGURE CAPTIONS}
\end{center}

\vspace{1cm}

\noindent
{\small FIG.~1: Diabatic potential surfaces for the reaction
  coordinate. An equilibrium wave packet is excited by a short laser
  pulse from a dark state (thin line) to the donor state (thick line,
  minimum at $-q_0$). From this nonequilibrium preparation a wave
  packet evolves on the coupled donor and acceptor (thick line,
  minimum at $+q_0$) surfaces where the transfer occurs in the
  Landau-Zener region near $q=0$.}

\vspace{0.3cm}

\noindent
{\small FIG.~2: Smoluchowski range for a harmonic RC according to
  (\ref{smolu2}). The dashed line separates the classical domain
  $\gamma\hbar\beta\ll 1$ from the quantum domain $\gamma\hbar\beta\gg
  1$.}

\vspace{0.3cm}

\noindent
{\small FIG.~3: Range of validity of the Quantum Zusman equations
  according to the first conditions in (\ref{val3}) (shaded) for fixed
  $E_{\rm r}/\hbar\Delta=2$. The
  dashed line separates the classical (small $\Delta\hbar\beta$) from
  the quantum range (larger $\Delta\hbar\beta$).}

\vspace{0.3cm}

\noindent
{\small FIG.~4: Quantum enhancement vs. inverse temperature for
  various damping strength together with the classical result
  $\kappa=1$.}

\vspace{0.3cm}

\noindent
{\small FIG.~5: Electron transfer rates vs. temperature for a symmetric system
  ($\epsilon=0$) according to the expression (\ref{rate4}) (solid
  line), the classical result ($\kappa=1$, dashed line),  and
  precise Quantum Monte Carlo data (diamonds, from
  \cite{lothar}). $E_{\rm r}/{\hbar\Delta}=10$ 
  and $\Delta/\Omega_c=2$.}

\vspace{0.3cm}

\noindent
{\small FIG.~6: Electron transfer rates vs. temperature for a symmetric system
  ($\epsilon=0$) according to the expression (\ref{rate4}) (solid
  line), the classical result ($\kappa=1$, dashed line), and
  exact Quantum Monte Carlo data (diamonds, from
  \cite{lothar}). $E_{\rm r}/{\hbar\Delta}=10$ 
  and $\Delta/\Omega_c=1$.}

\pagebreak

\begin{figure}
\center
\includegraphics[width=16cm,draft=false]{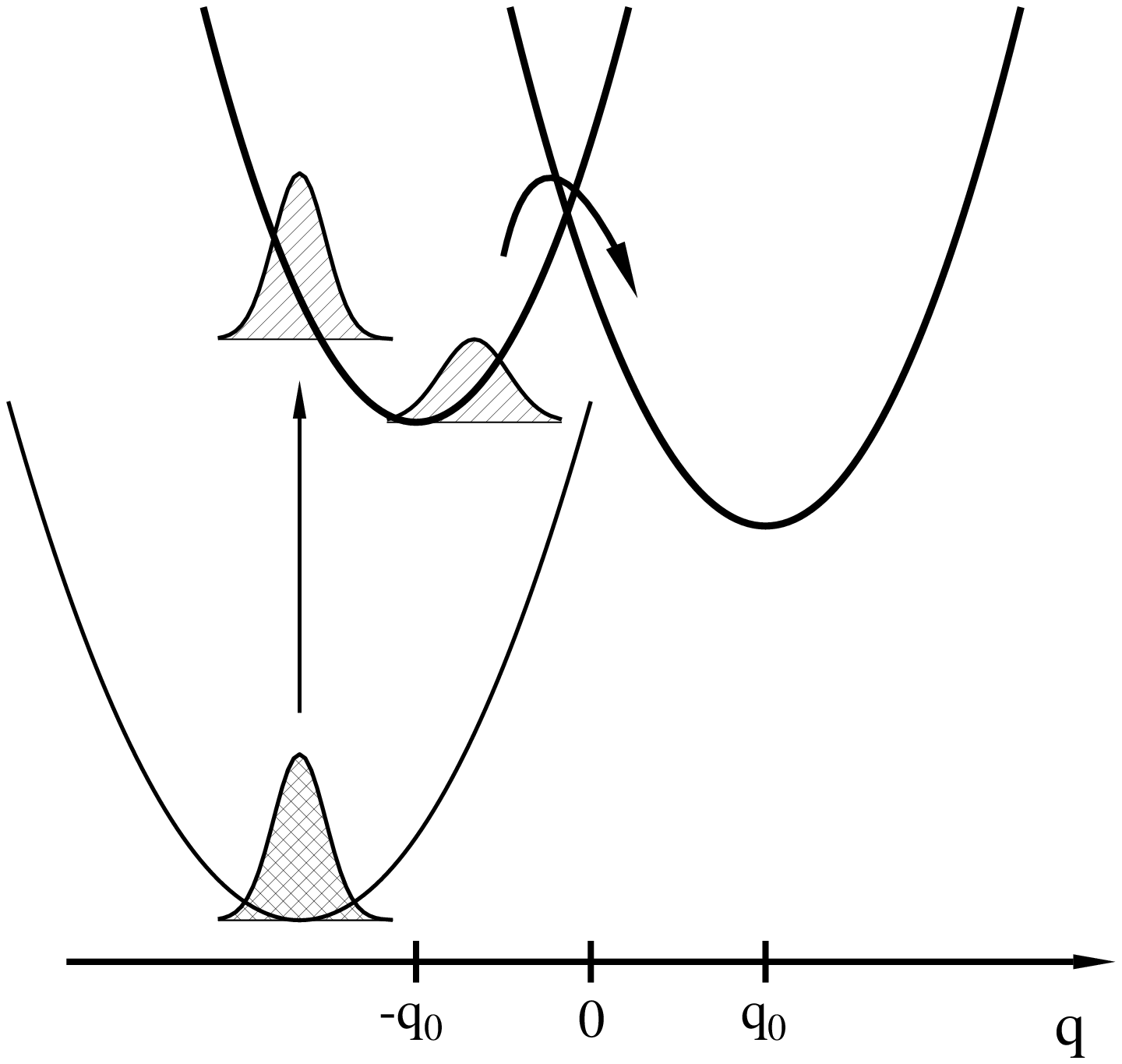}
%\caption{}
\label{fig1}
\end{figure}

\pagebreak

\begin{figure}
\center
\includegraphics[width=16cm,draft=false]{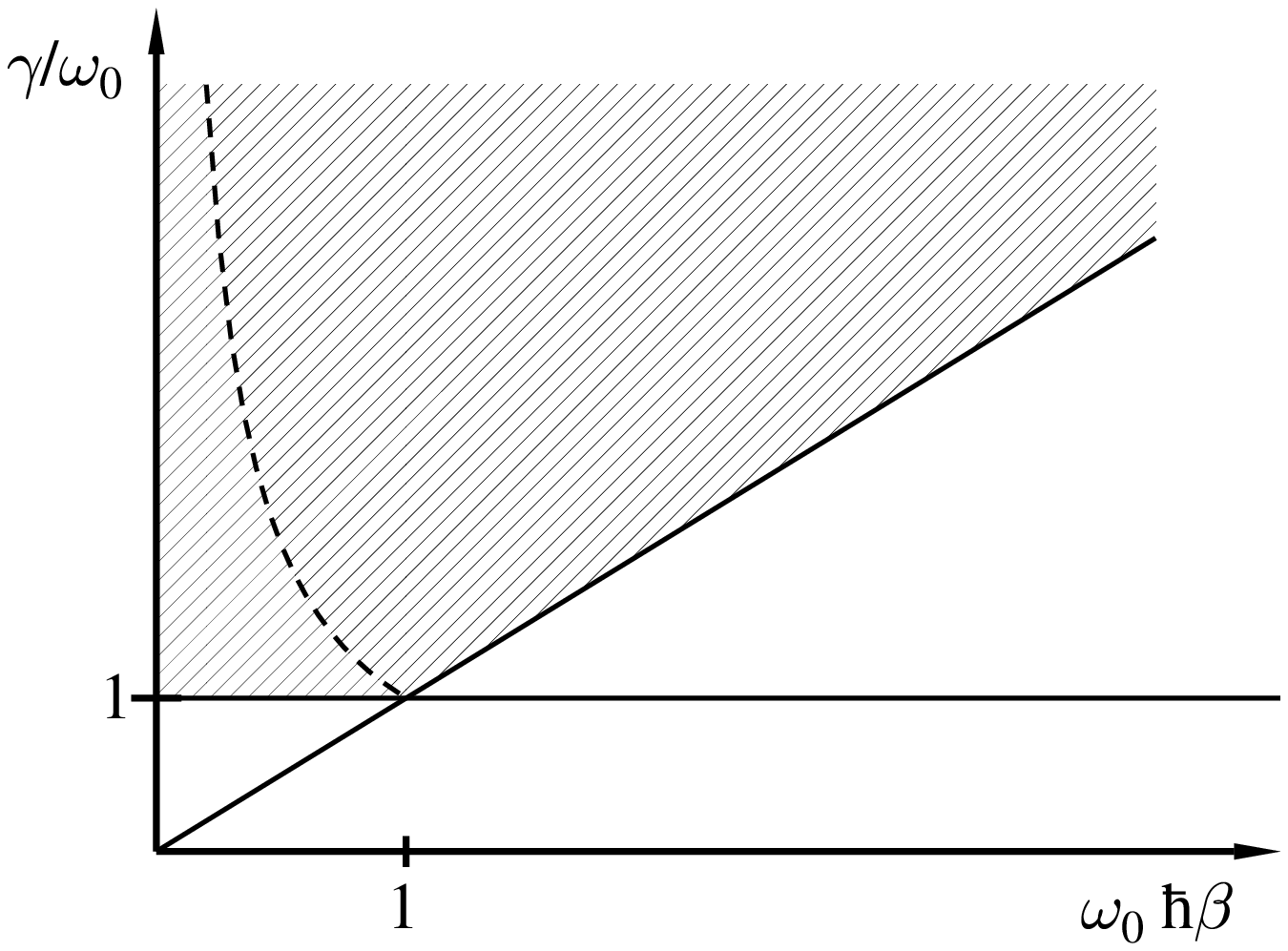}
%\caption{}
\label{fig2}
\end{figure}

\pagebreak

\begin{figure}
\center
\includegraphics[width=16cm,draft=false]{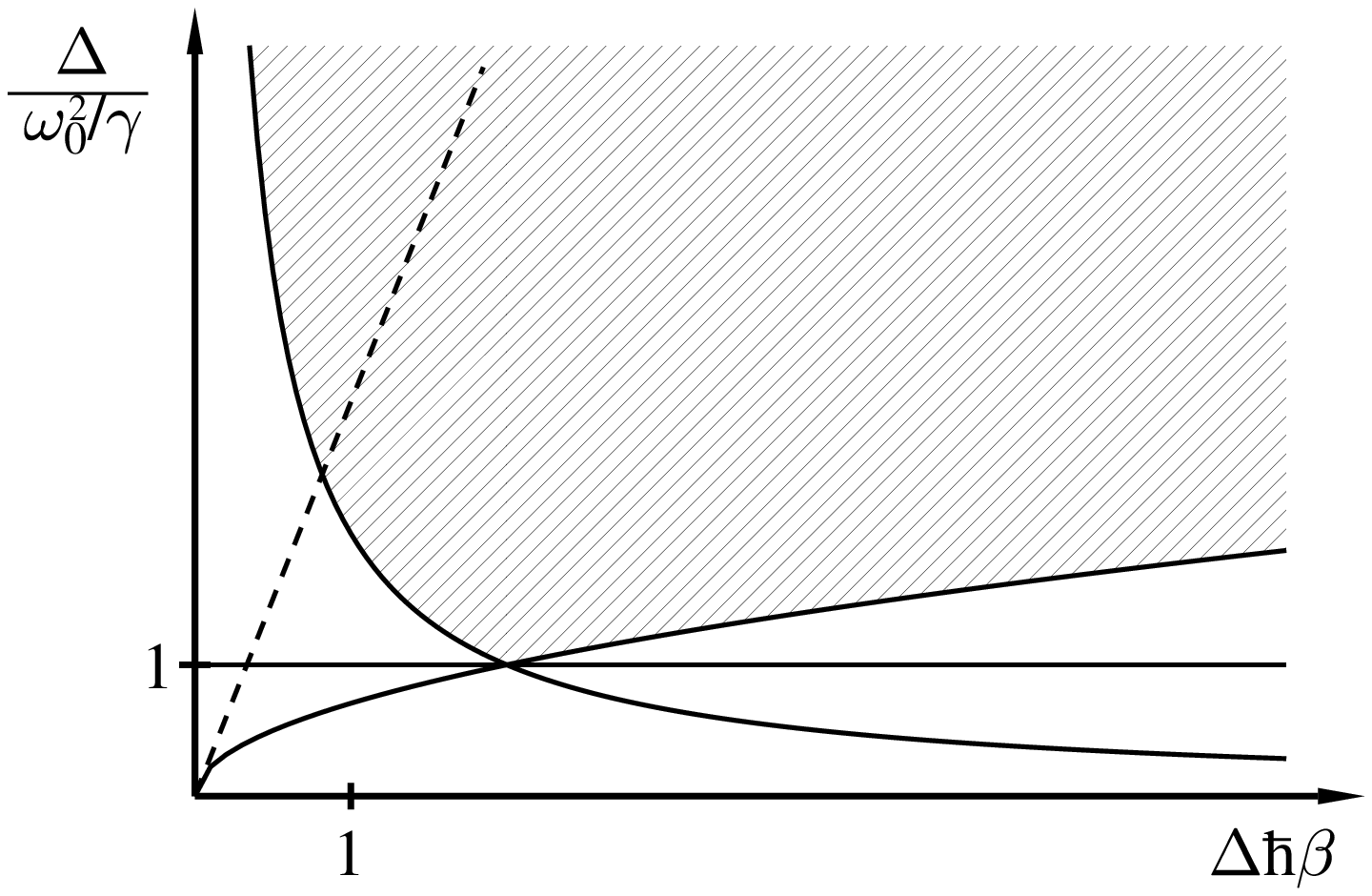}
%\caption{}
\label{fig3}
\end{figure}

\pagebreak

\begin{figure}
\center
\includegraphics[width=16cm,draft=false]{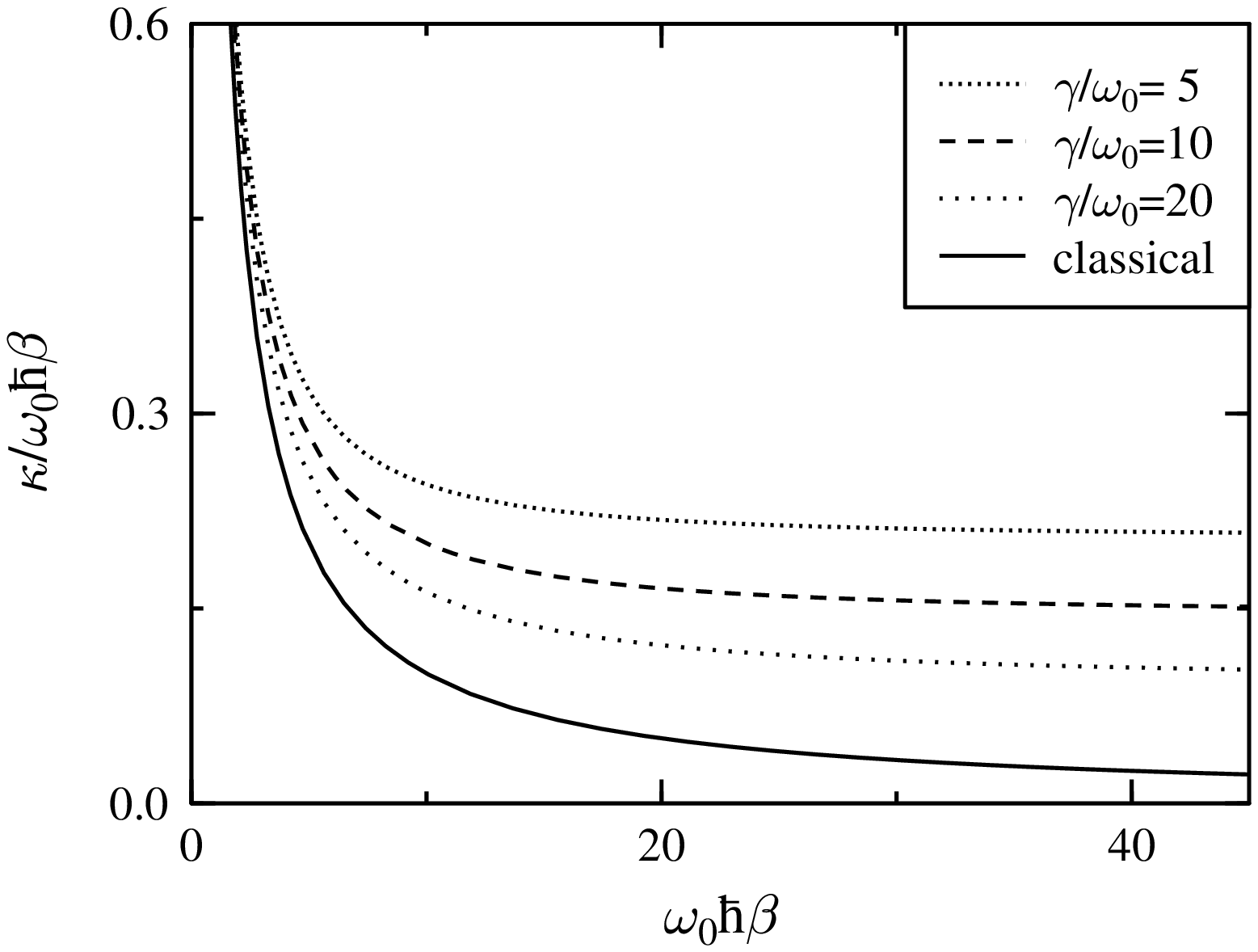}
%\caption{}
\label{fig4}
\end{figure}

\pagebreak

\begin{figure}
\center
\includegraphics[width=16cm,draft=false]{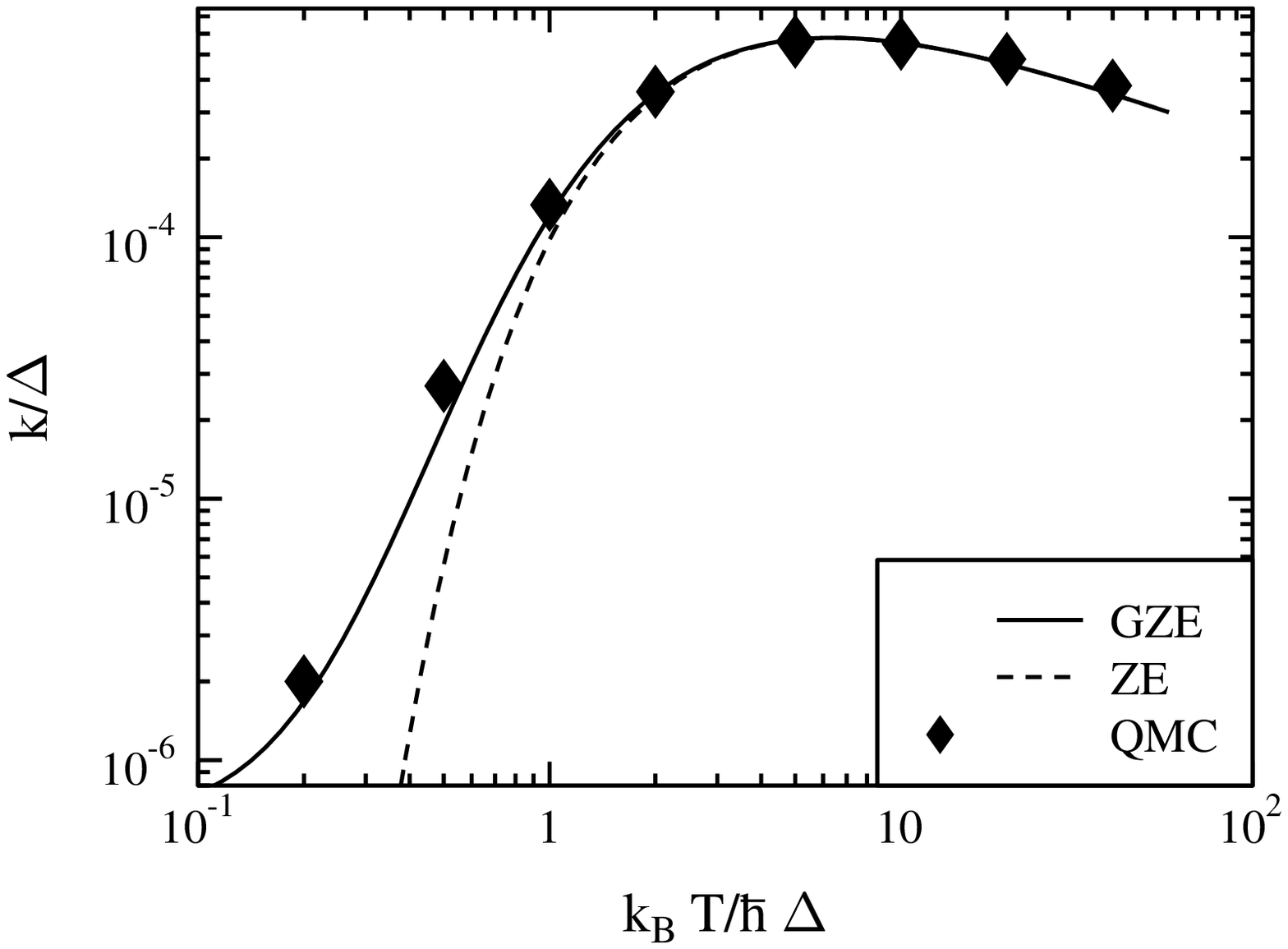}
%\caption{}
\label{fig5}
\end{figure}

\pagebreak

\begin{figure}
\center
\includegraphics[width=16cm,draft=false]{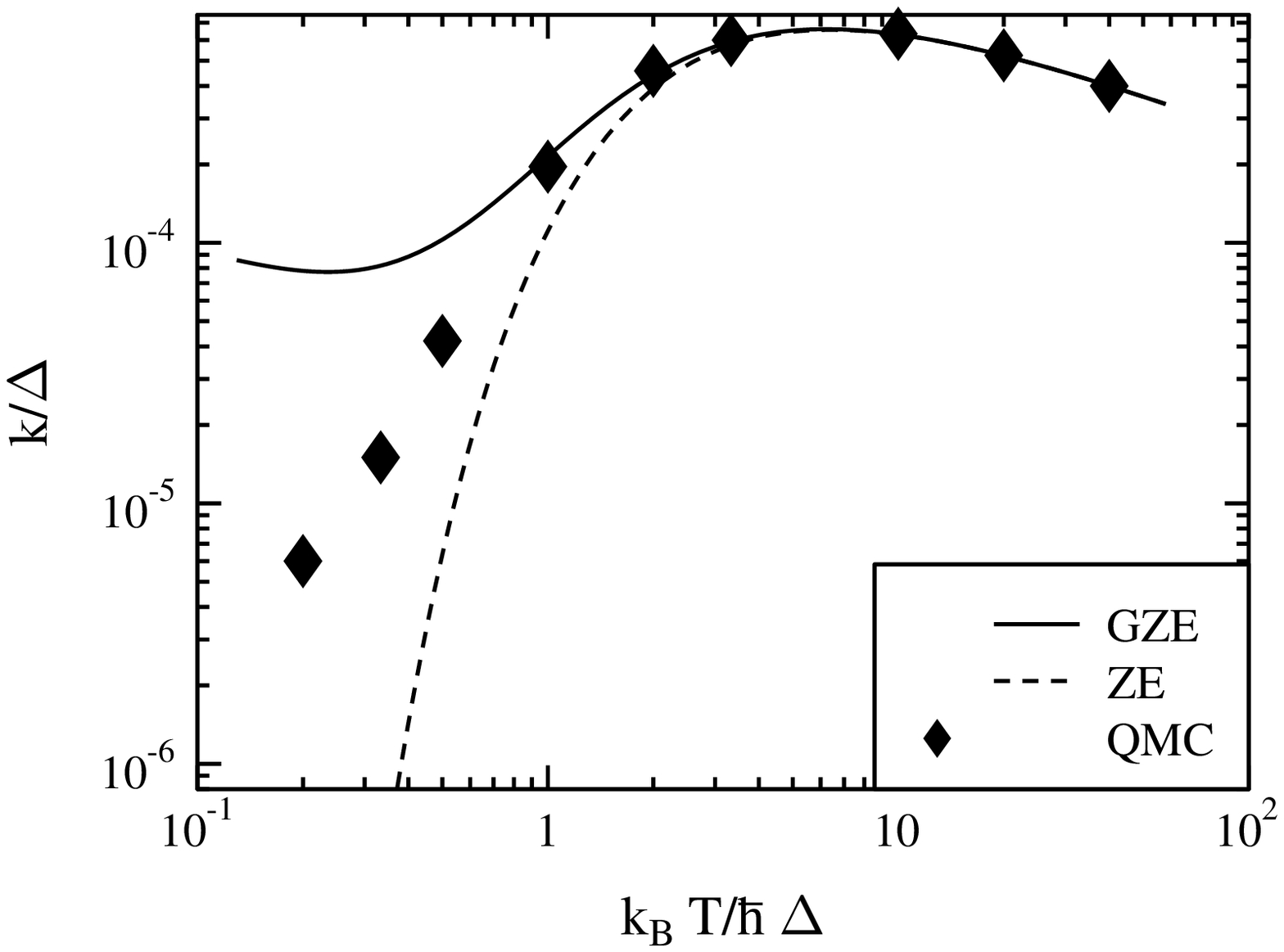}
%\caption{}
\label{fig6}
\end{figure}


\begin{thebibliography}{99}
\bibitem{rem} In this case $\gamma/\omega_0^2\approx 1$ps is the solvent
  autocorrelation time so that with $\hbar\beta\approx 10$fs (room
  temperature) follows $k_{\rm B} T
  (\gamma/\omega_0^2)/\hbar\beta\approx 1$ eV. This is of the same
  order of magnitude as the reorganization energy $E_{\rm r}$.
\end{thebibliography}
\end{document}